\documentclass[oneside,reqno,12pt,a4paper]{amsart}
\usepackage[T1]{fontenc}
\usepackage[margin=2.5cm]{geometry}
\usepackage{amsmath,amsfonts,amssymb,amscd,amsthm}
\usepackage[usenames,dvipsnames,svgnames]{xcolor}
\usepackage[utf8x]{inputenc}
\usepackage{graphicx}
\graphicspath{{}}
\usepackage{caption}
\usepackage{subcaption}
\usepackage{adjustbox}
\usepackage[numbers]{natbib}
\usepackage{doi}
\usepackage{makecell}
\usepackage{hyperref}
\usepackage{acronym}
\usepackage{listings}
\usepackage{textgreek}
\usepackage{url}
\usepackage{color}
\usepackage{framed}
\usepackage{verbatim}
\usepackage{fancyvrb}
\usepackage{newtxtext}
\usepackage{newtxmath}
\usepackage{microtype} 
\usepackage[final]{pdfpages}
\usepackage{tikz}
\usepackage{booktabs}
\usepackage{multirow}
\usepackage{algorithm,algorithmic}
\usepackage{mathtools}
\usepackage[normalem]{ulem}

\mathtoolsset{showonlyrefs}

\hypersetup{
breaklinks=true,
bookmarksopen=true,
pdftitle={Space-time unfitted finite elements on moving explicit geometry representations},    
pdfauthor={S. Badia, P. A. Martorell, F. Verdugo},     
colorlinks=true,       
linkcolor=black,          
citecolor=blue,        
filecolor=black,      
urlcolor=blue           
}

\definecolor{bg}{rgb}{0.93,0.93,0.93}

\acrodef{pde}[PDE]{partial differential equation}
\acrodef{fe}[FE]{finite element}
\acrodef{fem}[FEM]{finite element method}
\acrodef{dof}[DOF]{degree of freedom}
\acrodefplural{dof}[DOFs]{degrees of freedom}
\acrodef{agfe}[AgFE]{aggregated finite element}
\acrodef{agfem}[AgFEM]{aggregated finite element method}
\acrodef{sfem}[StFEM]{standard finite element method}
\acrodef{xfem}[XFEM]{extended finite element method}
\acrodef{cg}[CG]{continuous Galerkin}
\acrodef{dg}[DG]{discontinuous Galerkin}
\acrodef{ale}[ALE]{arbitrary Lagrangian Eulerian}
\acrodef{jit}[JIT]{just-in-time}
\acrodef{cad}[CAD]{computer-aided design}
\acrodef{fsi}[FSI]{fluid-structure interaction}
\acrodef{stl}[STL]{stereolithography}
\acrodef{amr}[AMR]{adaptive mesh refinement}
\acrodef{lic}[LIC]{line integral convolution}

\newcommand{\tnor}[1]{
  {\left\vert\kern-0.25ex\left\vert\kern-0.25ex\left\vert #1 
  \right\vert\kern-0.25ex\right\vert\kern-0.25ex\right\vert}}
\newcommand{\jnor}[1]{
  {\left[\kern-0.5ex\left[ #1 
 \right]\kern-0.5ex\right]}}

\newcommand{\fig}[1]{Fig.~\ref{#1}}
\newcommand{\sect}[1]{Sect.~\ref{#1}}

\newcommand{\alg}[1]{Alg.~\ref{#1}}

\newcommand{\alglin}[1]{line~\ref{#1}}

\definecolor{shadecolor}{gray}{.92}
\definecolor{incolor}{rgb}{0,0,.7}
\definecolor{outcolor}{rgb}{.65,0,0}
\definecolor{syntaxcolor}{rgb}{.65,0,0}

\newcommand{\julia}[1]{\mint{julia}|#1|}

\newif\ifsvgs
\svgsfalse

\newcommand{\includefig}[3][\tiny]{%
    \def\svgwidth{#2}
    #1
    
  \ifsvgs
    \updatepdffromsvg{#3}
  \fi
  \input{#3.pdf_tex}

}

\newcommand{\updatepdffromsvg}[1]{
  \executeiffilenewer{#1.svg}{#1.pdf}%
  {inkscape -z -C --file=#1.svg %
  --export-pdf=#1.pdf --export-latex && %
  sed "s|$(basename #1).pdf|$(echo #1).pdf|g" -i #1.pdf_tex }%
}

\newcommand{\executeiffilenewer}[3]{%
  \IfFileExists{#2}{}{\immediate\write18{#3}}
  \ifnum\pdfstrcmp{\pdffilemoddate{#1}}%
  {\pdffilemoddate{#2}}>0%
  {\immediate\write18{#3}}\fi%
}

\begin{document}
\title[Space-time unfitted FEM on explicit geometries]{Space-time unfitted finite elements on moving explicit geometry representations}
\author[S. Badia]{Santiago Badia$^{1,2}$}
\author[P. A. Martorell]{Pere A. Martorell$^{3,*}$}
\author[F. Verdugo]{Francesc Verdugo$^{4}$}
\thanks{\null\\
$^{1}$ School of Mathematics, Monash University, Clayton, Victoria, 3800, Australia.\\
$^{2}$ Centre Internacional de M\`etodes Num\`erics a l'Enginyeria, Campus Nord, 08034, Barcelona, Spain.\\
$^{3}$ Department of Civil and Environmental Engineering, Universitat Polit\`ecnica de Catalunya, Edifici C1, Campus Nord UPC,
Jordi Girona 1-3, 08034 Barcelona, Spain.\\
$^{4}$ Department of Computer Science, Vrije Universiteit Amsterdam, De Boelelaan 1111, 1081 HV Amsterdam, The Netherlands.
\\
$^*$ Corresponding author.\\
E-mails: {\tt santiago.badia@monash.edu} (SB),
{\tt pere.antoni.martorell@upc.edu} (PM),
{\tt f.verdugo.rojano@vu.nl} (FV)
}

\begin{abstract}

This work proposes a novel variational approximation of partial differential equations on moving geometries determined by explicit boundary representations. The benefits of the proposed formulation are the ability to handle large displacements of explicitly represented domain boundaries without generating body-fitted meshes and remeshing techniques. For the space discretization, we use a background mesh and an unfitted method that relies on {the} integration on cut cells. We perform this intersection by using clipping algorithms. To deal with the mesh movement, we pullback the equations to a reference configuration (the spatial mesh at the initial time slab times the time interval) that is constant in time. This way, the geometrical intersection algorithm is only required in 3D, another key property of the proposed scheme. At the end of the time slab, we compute the deformed mesh, intersect the deformed boundary with the background mesh, and consider an exact transfer {mechanism} between meshes to compute jump terms in the time discontinuous Galerkin integration. The transfer is also computed using geometrical intersection algorithms. We demonstrate the applicability of the method to fluid problems around rotating (2D and 3D) geometries described by oriented boundary meshes. We also provide a set of numerical experiments that show the optimal convergence of the method. 

\end{abstract}
\maketitle


\section{Introduction} \label{sec:intro}
Space-time formulations for \acp{fem} are valuable techniques for solving transient numerical problems. Unlike standard time stepping schemes, which employ different discretizations for space and time, space-time formulations discretize the problem simultaneously in both space and time. 
However, generating 4D space-time body-fitted meshes becomes extremely challenging on moving complex geometries. There are no general-purpose tools to generate 4D meshes. Besides, the re-meshing process due to moving domains introduces projection errors when transferring between meshes.

The bottleneck of mesh generation can be addressed by employing unfitted \acp{fem}. Unfitted \acp{fem}, also known as embedded or immersed \acp{fem}, {offer} a solution that eliminates the need for body-fitted mesh generation, relying instead on a simple background grid, such as a uniform or adaptive Cartesian grid. For large-scale simulations on distributed-memory platforms, one can replace expensive (both in terms of computational time and memory) unstructured mesh partitioning \cite{karypis_parmetis97} by efficient tree-mesh partitioners with load balancing \cite{burstedde_p4est_2011,Burstedde2016}. Unfitted \acp{fem} have gained increasing popularity in various applications, including \ac{fsi} \cite{Burman_2014a,Formaggia_2021,Schott_2019}, fracture mechanics \cite{Dekker_2019,Giovanardi_2017}, additive manufacturing  \cite{Carraturo_2020,Neiva_2020}, and stochastic geometry problems \cite{Badia_2021_uq}. Traditionally, unfitted problems utilize level sets to describe geometries. However, recent work \cite{Badia_2022-stl,Martorell2023} has extended unfitted discretizations to simulate geometries described by \ac{cad} models, i.e., explicit geometry representations.

The small cut cell problem, commonly discussed in the literature \cite{DePrenter2017}, is a significant limitation of unfitted \acp{fem}. The intersection between the physical domain and the background cells can become arbitrarily small, thereby giving rise to ill-conditioning issues. While several authors have attempted to address this problem, only a few techniques have demonstrated robustness and optimal convergence. The ghost penalty methods \cite{burman2010ghost}, employed within the so-called CutFEM framework \cite{burman_cutfem_2015}, represent one such approach. Alternatively, cell agglomeration techniques have emerged as viable options to ensure robustness in the presence of cut cells, naturally applied on \ac{dg} methods \cite{muller2017high}. Extensions to the $\mathcal{C}^0$ Lagrangian \ac{fe} have been introduced in \cite{Badia2018c}, while mixed methods have been explored in \cite{Badia2018a}, where the \ac{agfem} term was coined. \ac{agfem} exhibits good numerical qualities, including stability, bounds on condition numbers, optimal convergence, and continuity concerning data. Distributed implementations have been exploited in \cite{Verdugo2019, Badia_2021a}, while \ac{agfem} has also been extended to $h$-adaptive meshes \cite{Neiva2021} and higher-order \ac{fe} \cite{Badia_2022-highorder}. In \cite{Badia_2022-ghost}, a novel technique combining ghost penalty methods with \ac{agfem} was proposed, offering reduced sensitivity to stabilization parameters compared to standard ghost methods.

In the body-fitted case, one can consider variational space-time formulations \cite{Le_Beau_1993, Tezduyar_2006, Thompson_1996} to avoid the generation of 4D meshes.
In these formulations, a body-fitted mesh is extruded using a geometric mapping technique to represent the temporal evolution of the boundary displacements. However, when dealing with large displacements, the geometric mapping process becomes ill-posed, often necessitating re-meshing. Similar challenges arise in body-fitted \ac{ale} \cite{Donea_1982, Nobile1999} schemes, where frequent re-meshing becomes necessary due to large changes in topology. Consequently, none of these formulations are robust when confronted with large displacements. Furthermore, the re-meshing process introduces bottlenecks within the simulation procedure and incurs in projection errors when transferring between meshes.

Despite the notable advantages of unfitted \acp{fem}, its application in the space-time domain \cite{Badia_2023,Heimann2023} 
is currently limited to implicit level set geometrical representations (in which 4D geometrical treatment is attainable), 2D explicit representations (in which 3D geometrical tools, e.g., in \cite{Badia_2022-stl}, can readily be used for space-time) or piecewise constant approximations of the geometry in time \cite{Lehrenfeld2019}. This limitation stems from the complexity involved in developing 4D geometrical tools for unfitted methods {with explicit boundary representations}. 

The present work introduces a novel variational space-time formulation for unfitted \acp{fem} that eliminates the need for complex 4D geometrical algorithms. Instead, our formulation relies on time-extruded \ac{fe} spaces. Furthermore, we incorporate an \emph{exact} inter-slab transfer mechanism through an intersection algorithm, which removes projection errors between time slabs. The main contributions of our work are as follows: 
\begin{itemize}
  \item We develop space-time formulations for unfitted \acp{fem} on moving 3D geometries described explicitly. 
  \item We propose to pull back the problem to a reference configuration that is constant in each time slab to avoid 4D geometrical algorithms.
  \item We propose and implement an exact {time slab} {transfer mechanism} that relies on 3D intersection algorithms. 
  \item We compute the solution of transient problems on complex unfitted domains with large displacements.
\end{itemize}

The outline of this paper is as follows. In Section \ref{sec:method}, we introduce the geometry description and the space-time \ac{fe} spaces employed in this work. In Section \ref{sec:weak}, we present the variational formulation through a model problem. We also define the integration measures and the inter-slab transfer mechanism. In Section \ref{sec:intersection}, we describe the intersection algorithm for time slab transfer. Then, in Section \ref{sec:results}, we present numerical results for space and time convergence, condition number tests, and a numerical example of a moving domain. Finally, in Section \ref{sec:conclusions}, we present our conclusions and future work.

\section{Space time unfitted finite element method}\label{sec:method}
In this section, we introduce the geometry description by utilizing a geometrical map defined in a space-time \ac{fe} space. We also define the unfitted space-time \ac{fe} spaces
that are necessary to support our proposed formulation.

\subsection{Geometry description for moving domains}\label{sec:geometry}
Let us consider an initial Lipschitz domain $\Omega_0$ in a description suitable for unfitted \ac{fem}. Following the methodology described in \cite{Badia_2022-stl}, we assume that $\Omega_0$ is represented by a parameterized surface, e.g., a \ac{stl} mesh or a \ac{cad} model. For simplicity, we consider an oriented surface mesh $\mathcal{B}_{h}^0$ where the interior corresponds to the domain $\Omega_0$. Let $[0, T]$ be the time interval of interest.

We consider the coordinates of the mesh nodes of $\mathcal{B}_{h}$ to be time-depedendent.
Consequently, $\mathcal{B}_h(t)$  and its corresponding interior $\Omega(t)$. This variation of coordinates can be represented by a map $\pmb{D}: \mathcal{B}_h(0) \times [0,T] \rightarrow \mathcal{B}_h(t)$ such that $\pmb{D}(\cdot,0)$ is the identity.

\begin{figure}[http]
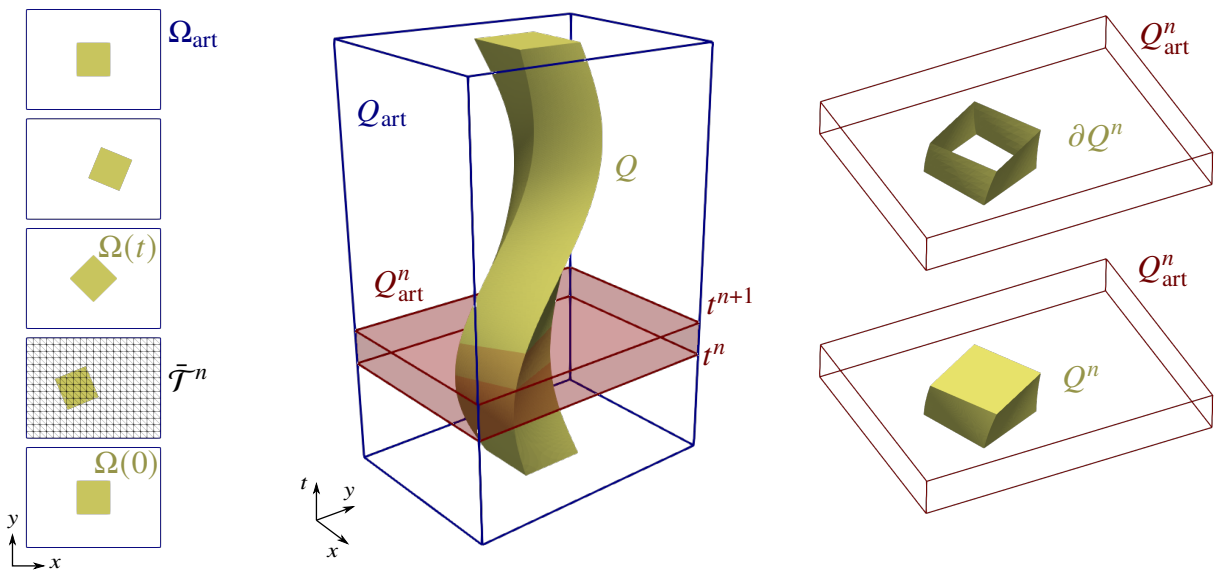

  \centering
  \includefig[\normalsize]{\textwidth}{notation_1}
  \caption[Representation of the space-time domain $Q$ embedded within an artificial space-time domain $Q_\mathrm{art}$.]{Representation of the space-time domain $Q$ embedded within an artificial space-time domain $Q_\mathrm{art}$. The spacial domains $\Omega(t)$ and $\Omega_\mathrm{art}$ in 2D are extruded in the time dimension. The solution is computed in each {time slab} $J^n=(t^{n},t^{n+1})$ on the space-time domain $Q^n$, which is embedded in $Q_\mathrm{art}$.}
  \label{fig:notation-domain}
\end{figure}

Next, we define the space-time domain $Q = \{ x\in \Omega (t) : t \in [0,T] \}$, in accordance with the nomenclature in \cite{Badia_2023}, see \fig{fig:notation-domain}. Additionally, we partition the boundary of the domain $\partial \Omega(t)$ into Dirichlet and Neumann boundaries,  $\partial \Omega_D(t)$ and $\partial \Omega_N(t)$, resp. This partition is such that 
$\partial \Omega(t) = \partial \Omega_D(t) \cup \partial \Omega_N(t)$ and $\partial \Omega_D(t) \cap \partial \Omega_N(t) = \emptyset$. The space-time boundaries are defined as  $\partial Q_* \doteq \bigcup _{t\in[0,T]} \partial \Omega _* (t) \times \{t\} $ for $* \in \{N,D\}$. Consequently, the boundary of $Q$ is given by $\partial Q \doteq \Omega (0) \cup \Omega (T) \cup \partial Q_D \cup \partial Q_N$.

In addition, we introduce an artificial spatial domain $\Omega_{\mathrm{art}}$, such that $\Omega(t) \subset \Omega_{\mathrm{art}}$ for all $t \in [0,T]$. This artificial domain can be a simple geometric shape, such as a bounding box, that can conveniently be meshed with a regular grid, such as a Cartesian mesh. Subsequently, we define the artificial space-time domain $Q_\mathrm{art} \doteq \Omega_{\mathrm{art}} \times [0,T]$, such that $Q \subset Q_\mathrm{art}$.

We define a partition in time, $\{J^n\}_{n=1}^N$ for $[0, T]$, in which the time domain is divided into time slabs. The time slabs are defined as $J^n = (t^{n},t^{n+1}), 1<n<N$ where $t^{n}<t^{n+1}, \forall n \in 1,...,N$. Within each time slab, the time-step size is defined as $\tau^n = t^{n+1} - t^{n}$ and the domain time-step size is defined as $\tau = \max _{n=1,..., N} \tau^n$. In each {time slab}, we define an artificial space-time domain  $Q^n_\mathrm{art} \doteq \Omega_\mathrm{art}\times J^n$. 
In a time slab, the space-time domain is determined by $Q^n \doteq   Q^n_\mathrm{art} \cap Q $.
Similarly, the space-time boundary is denoted as $\partial Q^n _{\{D, N\}}$.
We also introduce the notation $\Omega^n \doteq \Omega(t^n),\ n = 1,..., N+1$. 

\begin{figure}[http]
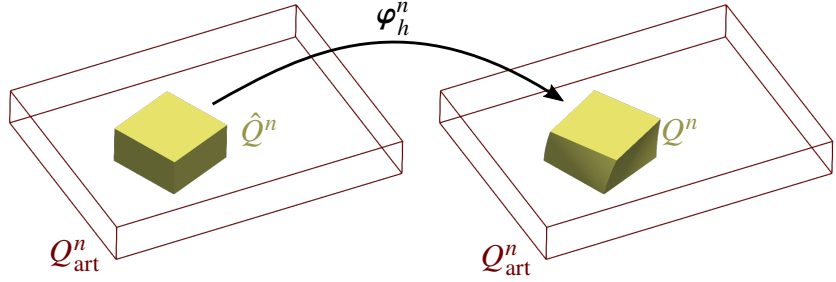

  \centering
  \includefig[\normalsize]{0.7\textwidth}{notation_2}
  \caption[Representation of the deformation map $\pmb{\varphi}^n_h$.]{ Representation of the deformation map $\pmb{\varphi}^n_h$ resulting from extending the variation of the surface map $\pmb D$ in the time slab domain $Q^n$ (or in the time slab artifical domain $Q^n_\mathrm{art}$). The undeformed configuration $\hat Q^n$ will be used for the \ac{fe} analysis. }
  \label{fig:notation-def}
\end{figure}

Finally, we define an undeformed space-time domain at each time slab as $\hat Q^n \doteq \Omega^{n} \times J^n$, see \fig{fig:notation-def}. 
Here, we note that $\Omega^{n}$ is the initial spatial domain of the time slab $J^n=(t^{n},t^{n+1})$. 
In order to define the reference configuration, we require a map $\pmb{\varphi}^{n}_{h}: \hat{{Q}}^n \rightarrow {Q}^{n}$ that must satisfy 
\begin{equation}\label{eq:map-req}
  \pmb{\varphi}^{n}_{h} (\partial \Omega(t)) = \pmb{D}(t) (\mathcal{B}_{h}^{0}),
\end{equation}
i.e., respect the boundary position determined by $\pmb{D}$.

Let $\bar{\mathcal{T}}_{\mathrm{art}}^n$  represent a simple partition of 
$\Omega_\mathrm{art}$ at time $t^{n}$.
Even though $\Omega_\mathrm{art}$ does not evolve in time, the partition $\bar{\mathcal{T}}_{\mathrm{art}}^n$ can differ across time slabs, e.g., when using \ac{amr} techniques.
We define the partition of 
$Q^n_\mathrm{art}$ as a Cartesian product of 
$\bar{\mathcal{T}}_{\mathrm{art}}^n$ and 
$J^n$. Specifically, $\mathcal{T}_{\mathrm{art}}^n \doteq \{ \bar K \times J^n : \bar K \in \bar{\mathcal{T}}^n_{\mathrm{art}}\}$.  
We can classify the spatial artificial cells $K\in \bar{\mathcal T}^n_\mathrm{art}$ as interior, cut, and exterior depending on the relative position concerning the domain boundary $\partial\Omega^{n}$. Since there is a one-to-one map between cells in $\bar{\mathcal T}^n_\mathrm{art}$ and ${\mathcal T}^n_\mathrm{art}$, it already provides a classification of the cells in the space-time artificial mesh $\mathcal{T}^n_\mathrm{art}$, which corresponds to the in-out classification for the extruded space-time domain boundary $\partial \hat{Q}^n$.

In the context of unfitted \ac{fem}, we are interested in solving \acp{pde} using the active portion of $\mathcal{T}_{\mathrm{art}}^n$, i.e., the cells touching {$\Omega^{n}$ at $t^{n}$.}
Thus, we remove the exterior cells $\bar{\mathcal{T}}^n_{\mathrm{out}} \doteq \{ \bar{K}\in \bar{\mathcal{T}}^n_{\mathrm{art}}: \bar{K}\cap\Omega^{n} = \emptyset \}$ from the artificial spatial partition $\bar{\mathcal{T}}^n_\mathrm{art}$, leading to an active spatial partition 
$\bar{\mathcal T}^n \doteq \bar{\mathcal{T}}^n_\mathrm{art} \setminus \bar{\mathcal{T}}^n_\mathrm{out}$.
From the spatial active partition, we compute the active space-time partition $\mathcal{T}_{}^n \doteq \{ \bar K \times J^n: \bar K \in \bar{\mathcal{T}}^n_{}\}$ by extrusion.

\subsection{Space-time finite element spaces}\label{sec:fem}

In this section, we define space-time unfitted \ac{fe} spaces for the discretization of the model problem on moving domains. However, a main difference with respect to \cite{Badia_2023}, the original problem on $Q^n$ will be re-state in the undeformed domain $\hat Q^n$. As a result, in this section, we only need a construction that works on time-extruded domains.  Thus, the active mesh $\mathcal T^n$ is the one that intersects $\Omega^{n}$ at $t^{n}$, as defined above. We note that $\mathcal{T}^n$ may change across time slabs in evolving domains. 

Firstly, we define the \ac{fe} space $\mathcal{X}^n_h$ in the spatial domain. Then, for each \ac{dof} in $\mathcal{X}_h$, we introduce a 1D \ac{fe} basis $\mathcal{Y}^n_\tau$. Thus, we construct the space-time \ac{fe} space composed as the tensor product of these two spaces, $\mathcal{V}^n_h \doteq \mathcal{X}^n_h \otimes \mathcal{Y}^n_\tau$.

In each space-time cell $T^n = \bar{T}^n \times J^n \in \mathcal{T}^n$,  we can define the \ac{fe} local interpolation, which consists of tensor product \acp{dof} $\Sigma \doteq \Sigma_X \otimes \Sigma_Y$  and shape functions $\Phi \doteq \Phi_X \otimes \Phi_Y$. Specifically, any shape function $\phi \in \Phi$ can be expressed as $\phi^{(\alpha_X,\alpha_Y)}(\pmb{x},t) = \phi^{\alpha_X}_X(\pmb{x}) \otimes \phi^{\alpha_Y}_Y(t)$. 

To address the ill-conditioning issues in unfitted \ac{fem}, we will utilize as a model example the \ac{agfem}, even though the proposed numerical framework can readily be used for other techniques, e.g., ghost penalty stabilization (see \cite{Badia_2023} for space-time unfitted formulations). The \ac{agfem} eliminates problematic \acp{dof} by constraining them to well-posed \acp{dof} using a discrete extension operator $\mathcal{E}$. We define the spatial extension operator $\bar{\mathcal{E}}: \bar{\mathcal V}_{h,\mathrm{in}} \mapsto \bar{\mathcal V}_h $ between the the spatial interior $\bar{\mathcal V}_{h,\mathrm{in}}$ and active $\bar{\mathcal V}_h$ \ac{fe} spaces . The spatial \ac{agfe} space is defined as $\bar{\mathcal V}_{h,\mathrm{ag}} \doteq \bar{\mathcal{E}} ( \bar{\mathcal V}_{h,\mathrm{in}})$

We define a slab-wise space-time discrete operator between $\mathcal V^n_{h,\mathrm{in}}$ and $\mathcal V^n_h$. In each time slab, we aggregate the cut cells in the spatial active mesh $\bar{\mathcal{T}}^n$ using the aggregation techniques from the \ac{agfem}.
For a given time slab $J^n$, the space-time extension operator $\mathcal{E}^n : \mathcal V^n_{h,\mathrm{in}} \mapsto \mathcal V^n_h$ is defined as
\begin{equation}
  \mathcal{E}^n(\mathcal V^n_{h,\mathrm{in}}) = \mathcal{E}^n(\bar{\mathcal V}^n_{h,\mathrm{in}}\otimes \mathcal{Y}^n_\tau) = \bar{\mathcal{E}}^n(\bar{\mathcal V}^n_{h,\mathrm{in}}) \otimes \mathcal{Y}^n_\tau,
\end{equation}

where $\bar{\mathcal{E}}^n: \bar{\mathcal V}^n_{h,\mathrm{in}} \mapsto \bar{\mathcal V}^n_h$ represents the spatial extension operator, and $\mathcal{Y}_\tau^n$ the \ac{fe} space in time at the time slab $J^n$. We define the space-time \ac{agfe} space on the {time slab}. Now, we can define the global \ac{agfe} space as $\mathcal V_{h,\mathrm{ag}} \doteq \mathcal V_{h,\mathrm{ag}}^1 \times \dots \times \mathcal V_{h,\mathrm{ag}}^N$. Note that no continuity is imposed across time slabs, i.e., the discrete time space is discontinuous.

In ghost penalty formulations, one can readily use the non-aggregated spaces on the active mesh and add stabilization terms (see, e.g., \cite{dePrenter2023}) to make the problem robust with respect to cut locations. Thus, the space-time \ac{fe} spaces to be used in this case is simply $\mathcal{V}_{h}^n$. 

\subsection{Extension of the deformation map}\label{sec:map-extension}
\begin{figure}[http]
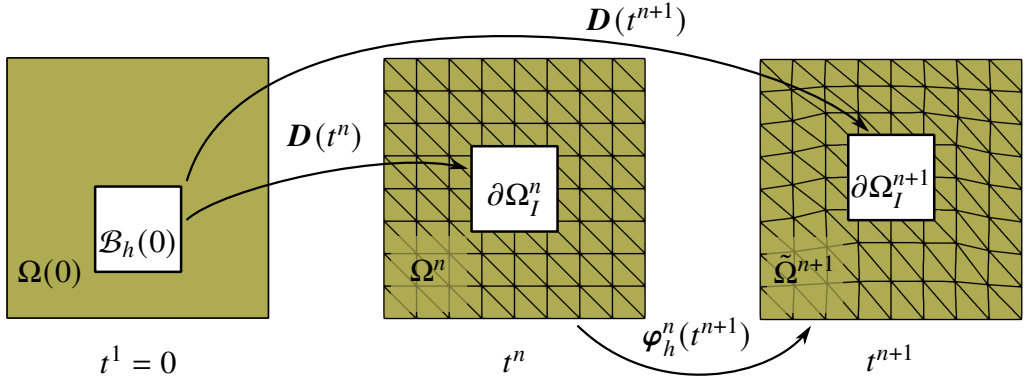

  \centering
  \includefig[\normalsize]{0.85\textwidth}{deformation_maps}
  \caption[Representation of the deformation map $\pmb{D}$ and the extended map $\pmb{\varphi}_h^n$ in the time slab $J^n=(t^{n},t^{n+1})$.]{Representation of the deformation map $\pmb{D}$ and the extended map $\pmb{\varphi}_h^n$ in the time slab $J^n=(t^{n-1},t^n)$. This example represents the external domain. The spatial deformation map $\pmb{D}(t): \mathcal B_h(0) \mapsto \mathcal B_h(t),\ t\in[0,T]$, is defined on the surface mesh, and the extended map $\pmb{\varphi}_h^n(t): \Omega^{n} \mapsto \tilde \Omega(t),\ t\in J^n$, is defined on the spatial domain. At $t^{n+1}$, the description of the inner boundary $\partial \Omega_I^{n+1}$ may differ across maps:
  $\pmb{\varphi}_h^n(t^{n+1})(\partial\Omega_I^{n}) \approx \pmb{D}(t^{n+1})( \mathcal B _{h}(0) )$, where $\partial\Omega_I^{n} =  \pmb{D}(t^{n})( \mathcal{B}_h(0) ) $. The approximation error decreases with the spatial discretization size $h$.}
  \label{fig:def-maps}
\end{figure}

In the case in which the domain is represented in time by its explicit boundary representation and its boundary displacement $\pmb{D}(t)$, we must extend $\pmb{D}(t)$ defined on $\mathcal B(t)$ to the space-time domain $\hat Q^n$ of each time slab $J^n$.
For this purpose, we solve the following linear elasticity problem in $\hat Q^n$, even though other extension operators could be considered. Find $\hat{\pmb{u}}\in \mathbb{R}^d$ such that, 
\[ 
  \begin{cases}
    \hat{\boldsymbol{\nabla}}_x \cdot \pmb{\sigma}(\hat{\pmb u}) = 0 & \mathrm{in}\ \hat Q^n,\\
    \hat{\pmb u} = \hat{\pmb u}_D^n & \mathrm{on}\ \partial \hat Q^n_D,\\
    \hat{\pmb u} = 0 & \mathrm{on}\ \partial \hat Q^n_{D_0}, \\
    \mathbf{n}_x \cdot \pmb{\sigma}(\hat{\pmb u})  = 0 & \mathrm{on}\ \partial \hat Q^n_N.
  \end{cases}
\]
where $\pmb{\sigma}$ is the stress tensor, {$\hat{\boldsymbol{\nabla}} _x$ is the spatial gradient, $\mathbf{n}_x$ is the spatial component of outward normal vector to  $\partial \hat Q^n$, and}
\[ \hat{\pmb{u}}_D^n(t) = \left( \pmb{D}(t) - \pmb{D}(t^{n}) \right)\circ \left( \pmb{D}(t^{n}) \right) ^{-1}  \]
is the Dirichlet boundary condition on $\partial \hat Q^n_D = \mathcal B_h(t^{n})\times J^n$,  which imposes the deformation near the unfitted boundary (see \fig{fig:def-maps}). On the boundary of the artificial domain and at the initial spatial domain,  $\partial \hat Q^n_{D_0} = \partial \Omega_\mathrm{art} \times J^n \cup \Omega^{n}\times \{t^{n}\}$, the deformation is fixed to zero. The Neumann boundary is defined on  
$\partial \hat Q^n_N = \Omega^n\times \{ t^{n+1}\}$.

This problem is approximated with a \ac{cg} method and weak imposition of the Dirichlet boundary conditions on the unfitted boundary $\partial \hat Q_D$ using Nitsche's method \cite{Nitsche1971}. In the weak formulation, we find $u\in \mathcal{V}^n_{h,\mathrm{ag}}$ such that,
\[ a(\hat{\pmb{u}},\hat{\pmb{v}}) = l(\hat{\pmb{v}}), \quad \forall \hat{\pmb{v}} \in \mathcal{V}^n_{h,\mathrm{ag}} \]
where the bilinear form and the linear form are defined as,
\[
a(\hat{\pmb{u}},\hat{\pmb{v}}) = 
\int_{\hat Q^n} \pmb{\varepsilon}(\hat{\pmb{v}}) : \pmb{\sigma}(\pmb{\varepsilon}(\hat{\pmb{u}})) d\hat{\pmb{x}}dt +
\int_{\partial \hat Q_D} \left( \tau_D \hat{\pmb{v}} \cdot \hat{\pmb{u}} - 
  v\cdot \left(\mathbf{n} _x \cdot \pmb{\sigma}(\pmb{\varepsilon}(\hat{\pmb{u}})) \right) \right) d\hat{\pmb{x}}dt,
\]
and
\[l(\hat{\pmb{v}}) = \int_{\partial \hat Q_D} \left( \hat{\pmb{v}} \cdot \hat{\pmb{u}}_D -
  \mathbf{n}_x \cdot \pmb{\sigma}(\pmb{\varepsilon}(\hat{\pmb{v}})) \cdot \hat{\pmb{u}}_D \right) d\hat{\pmb{x}}dt,\]
resp., where $\pmb{\varepsilon}$ is the symmetric (spatial) gradient operator. 

From the approximated solution $\hat{\pmb{u}}^n_h \in \mathcal{V}^n_{h,\mathrm{ag}}$ we extract the map 
$\pmb{\varphi}^n_h(\hat{\pmb{x}},t) = \hat{\pmb{x}} + \hat{\pmb{u}}^n_h(\hat{\pmb{x}},t)$. One can alternatively use a ghost penalty stabilization to solve the elasticity problem. Since the deformation is imposed weakly, the deformation $\hat{\pmb{u}}^n_h$ is not equal to $\hat{\pmb{u}}_D^n$ on $\mathcal B_h(t)$.
Thus, the deformed space-time domain $Q^n$ is approximated by $\tilde Q^n = \pmb{\varphi}^n_h ( \hat Q^n)$.
In any case, the geometrical error is expected to decrease with the mesh size $h$. 
For given time $t\in J^n$, we can extract the spatial map $\pmb{\varphi}^n_h(t): \Omega^{n} \mapsto \tilde \Omega(t)$ (see \fig{fig:def-maps}). Here, the spatial domain is also approximated, $\Omega(t) \approx \tilde \Omega(t)$. 

We assume that the computed map is one-to-one, i.e., $\mathrm{det}(\boldsymbol{\nabla}\pmb{\varphi}_h^n) > 0$ at all times. Assuming that $\pmb{D}$ is such that it admits an extension that is diffeomorphic, this requirement can be attained for small enough time steps in the time-discrete problem. One can also move back to the Cartesian mesh after several time steps, as soon as the map remains bijective.

\subsection{Extended active mesh}

As discussed above, in the case in which the deformation map has to be computed, $\pmb{\varphi}^n_h$ is not equal to $\pmb{D}$ on $\mathcal{B}_h(t)$, since unfitted methods usually make use of weak imposition of boundary conditions. 

At the end of the time slab $J^n$, we can define the solution in the approximated domain $\tilde{Q}^n$ as
\[
{u}_h^{n+1,-}(\pmb{x}) = \hat{u}_h^n( \pmb{\varphi}^n_h(t^{n+1})^{-1}(\pmb{x}),t^{n+1}), \quad \forall \pmb{x} \in \tilde{\Omega}^{n+1},
\] 
which is a function defined on the undeformed domain $\Omega^{n}$; since we integrate the forms on $\hat{Q}^n$, the inverse map is never computed in practice. 
On the other side, with the method proposed below, we have to compute an inter-slab integral on $\Omega^{n+1}$ that involves $u^{n+1,-}_{h}$.

In order for $u_{h}^{n+1,-}$ to be defined on $\Omega^{n+1}$, we proceed as follows. First, we extend the active mesh by $\bar{\mathcal{T}}_\mathrm{ext}^n = \bar{\mathcal T}^n \cup \{ K \in \bar{\mathcal{T}}^n_\mathrm{art}: \pmb{\varphi}^n(t^{n+1}) ( K ) \cap \Omega^{n+1}  \ne \emptyset\}$. Assuming that  the geometrical map is one-to-one, the inverse $\pmb{\varphi}^n_h(t^{n+1})^{-1}$ is well-defined on $\Omega^{n+1}$.   

To accommodate a \ac{fe} space $\bar{\mathcal V}_{h,\mathrm{ext}}^n$ in $\bar{\mathcal T}_\mathrm{ext}^n$, one can consider the modification of the extension operator.
In the \ac{agfem} framework, we can simply redefine the extension operator $\bar{\mathcal{E}}^n_\mathrm{ext} : \bar{\mathcal V}^n_{h,\mathrm{in}} \mapsto \bar{\mathcal V}_{h,\mathrm{ext}}^n$. 
In non-aggregated methods like CutFEM, we can utilize the same extension on the solution $\bar{\mathcal{E}}^n_\mathrm{ext} (\bar{\mathcal V}^n_h)$.

We note that the implementation of the extended triangulation and \ac{fe} space can be computed a posteriori, on demand, whenever it is needed. One can mark the additional cells that are needed for the extension, and then extend the aggregates to these cells.

\section{Variational formulation on a model problem}\label{sec:weak}

In this section, we establish the space-time variational formulation by employing a model problem, specifically the heat equation. Although we use the heat equation for demonstration purposes, it is essential to note that a similar approach can be applied to other \acp{pde}.

\subsection{Weak formulation}\label{sec:weak-form}
In order to define the space-time variational formulation, we first define the convection-diffusion equation in a space-time domain $Q$, as find $u$ such that,
\begin{equation}\label{eq:heat}
  \begin{cases}
  \partial_t u + (\pmb{w}\cdot \boldsymbol{\nabla}_x) u 
  - \mu\Delta_x u = f \quad &\text{in } Q, \\
  u = u_D \quad &\text{on } \partial Q _D,\\
  \mu \mathbf{n}_x \cdot \boldsymbol{\nabla}_x u  = g_N \quad &\text{on } \partial Q _N, \\
  u = u_0 \quad &\text{on } Q(0).
  \end{cases}
\end{equation}
{where $\mu$ is the diffusion coefficient, $\pmb{w}$ the advection velocity, $f$ the source term, $u_D$ the Dirichlet boundary condition, and $g_N$ boundary flux on $\partial Q_N$. $\boldsymbol{\nabla}_x$ and $\Delta_x$ denote the spatial gradient and spatial Laplacian, resp.}
Well-posedness requires that {$\pmb{w}\cdot \mathbf{n}_x + n_t \geq 0$} on the Neumann boundary $\partial Q _N$. { Here, $\mathbf{n} = (\mathbf{n}_x,n_t)$ is the outward normal to $\partial Q$.}

We discretize this problem with the \ac{agfem} in space (or a ghost penalty stabilization ) and a \ac{dg} method in time. We weakly impose the Dirichlet boundary conditions using Nistche's method \cite{Nitsche1971}. 
Since the coupling between time slabs respects causality, we analyze the problem on a single time slab, assuming we know the solution of the previous one (see also \cite{Badia_2023}).
To analyze each time slab $J^n=(t^{n},t^{n-1})$, we define the problem in the reference domain  $\hat Q^n = (\pmb{\varphi}^n_h)^{-1} ( \tilde Q^n) $ as: find $\hat u \in \mathcal{V}^n_{h,\mathrm{ag}}$ such that
\begin{equation}
  B_h^n(\hat u, \hat v) =L_h^n(\hat v),\quad \forall \hat v \in \mathcal{V}^n_{h,\mathrm{ag}},
\end{equation}
with $u =\hat u \circ (\pmb{\varphi}^n_h)^{-1}$. The bilinear form reads as: 
\begin{equation}\label{eq:bilinear-form}
 \begin{aligned}
  B_h^n(\hat u, \hat v) &= 
   \int_{\hat Q^n}  \hat v \partial_t^n \hat u  | J  _ {\tilde{Q}^n}| d\hat{\pmb{x}} dt
   + \int_{\Omega^{n}} 
    \hat v (t^{n}) \hat u (t^{n})
    | J _{\Omega^{n}}|d\hat{\pmb{x}}
    + a_h(\hat u, \hat v),\\
  a_h(\hat u, \hat v) &=
    \int_{\hat Q^n} \left( \mu  \boldsymbol{\nabla} ^n_x \hat u \cdot \boldsymbol{\nabla}^n _x \hat v \ + \  
     \hat v (\pmb{w} \circ \pmb{\varphi}_h^n \cdot  \boldsymbol{\nabla} ^n_x)  \hat u  \right) | J  _ {\tilde{Q}^n}| d\hat{\pmb{x}} dt\\
   &+ \int_{\partial \hat Q^n_D}  \left( 
    \beta_h  \hat v \hat u 
    - \hat v \left( \mathbf{n}^n_{x} \cdot \mu\boldsymbol{\nabla}^n_x \hat u \right)
    - \left( \mathbf{n}^n_{x} \cdot \mu\boldsymbol{\nabla}^n_x \hat v \right) \hat u
    \right) |J_{\partial \tilde{Q}^n_D} | d\hat{\pmb{x}} dt,\\
\end{aligned}
\end{equation}
and the linear form reads as,
\begin{equation}\label{eq:linear-form}
\begin{aligned}
  L^n_h(\hat v) &= 
  \int_{\Omega^{n}}  
  \hat v (t^{n})   \hat u ^{n-1}(t^{n})
  | J _{\Omega^{n}}|d\hat{\pmb{x}}
  + l_h(\hat v) , \\
  l_h(\hat v) &=
  \int_{\hat Q^n} \hat v  \left( f \circ \pmb{\varphi}_h^n\right) | J  _ {\tilde{Q}^n}| d\hat{\pmb{x}} dt 
  + \int_{\partial \hat Q^n_N}  \hat v \left( g_N \circ \pmb{\varphi}_h^n\right) |J_{\partial \tilde{Q}^n_N} | d\hat{\pmb{x}} dt
 \\
   &+ \int_{\partial \hat Q^n_D}  \left( 
    \beta _ h \hat v \left(  u_D \circ \pmb{\varphi}_h^n  \right)
    - \left( \mathbf{n}^n_{x} \cdot \mu\boldsymbol{\nabla}^n_x \hat v \right) 
    \left(  u_D \circ \pmb{\varphi}_h^n  \right)
    \right) |J_{\partial \tilde{Q}^n_D} | d\hat{\pmb{x}} dt, 
\end{aligned}
\end{equation}
Let us define the norms used in \cite{Badia_2023} to prove stability and convergence results, which we will computed in the numerical experiments. In \cite{Badia_2023} the space-time accumulated \ac{dg} norm of $\mathcal V_{h,\mathrm{ag}}$ is defined as follows,
\begin{equation}\label{eq:dg-norm}
  \tnor{v}^2_{n,*}  \doteq \|v^{n}(t^{n+1}) \|^2_{L^2(\Omega^n)} + \sum^{n}_{i=1}\| v^{i}(t^{i}) - v^{i-1}(t^{i}) \|^2_{L^2(\Omega^i)} + c_\mu \int^{t^{n+1}}_0 \| v \|^2 _{\bar{\mathcal{V}} ^n(h)} dt,
\end{equation}
where $c_\mu$ is the coercivity constant and $\bar{\mathcal{V}}^n(h) \doteq \bar{\mathcal{V}}^n_{h,\mathrm{ag}}+H^2(\Omega(t))$ is the norm of the \ac{fe} space at a time step $t$ given by
\begin{equation}
  \tnor{v}^2 _{\bar{\mathcal{V}} ^n(h)} \doteq \mu \| \boldsymbol{\nabla} v\| ^2 _{L^2(\Omega(t))} +
  \sum_{\bar{T}\in\bar{\mathcal{T}}_{h,\mathrm{act}}} \beta_{\bar{T}} \|v\|^2_{L^2( \bar{T} \cap \partial \Omega _D(t)) }
  + \sum_{\bar{T}\in\bar{\mathcal{T}}^n_{h,\mathrm{act}}} \mu h^2_{\bar{T}} \|v\|^2_{H^2( \bar{T} \cap  \Omega(t) ) }.
\end{equation}

The proof of this stability result follows the ideas in \cite{Badia_2023} in the case in which the deformation map $\pmb{\varphi}_h^n(t)$ is equal to $\pmb{D}(t)$ on $\Gamma(t)$. Otherwise, the analysis is more technical and would require to use ideas similar to the ones in \cite{Lehrenfeld2019,heimann2023geometrically}. The analysis therein assumes a constant deformation map and a level-set description of the domain. In our case, the deformation map can be of higher-order (time-dependent) and the domain is represented by its boundary. We will leave this analysis for future work.

In the variational form, the Dirichlet boundary conditions are imposed weakly with the Nistche method, with a penalty term $\beta_h$ that depends on the spatial cell size $h$. The initial value is also imposed weakly with \ac{dg} in time, where the jump is given by $ [\hat u (t^{n}) - \hat u^{n-1}(t^{n}) ]$. The evaluation of the solution of the previous time slab $\hat u^{n-1}(t^{n}) $ in $\Omega^{n}$
requires special attention since it is computed using a different discretization. Further discussion is provided in \sect{sec:inter-slab} and \sect{sec:intersection}.

The derivatives are projections of the space-time gradient into space and time, 
$\partial ^n_t = ( \boldsymbol{\nabla}^n )_t,$ 
and $ \boldsymbol{\nabla} ^n _x = (  \boldsymbol{\nabla} ^n )_x,$ resp. They are obtained by transporting the space-time gradient to the deformed domain
$ \boldsymbol{\nabla} ^n = \mathbf F^{-T} \hat{\boldsymbol{\nabla}},$ where $\mathbf F = \boldsymbol{\nabla} \pmb{\varphi}_h^n$. The operators $(~)_x$ and $(~)_t$ represent the projection to the space and time directions, resp.
The boundary normal is also transported to the deformed domain. It is computed as follows,
\begin{equation}
  \mathbf{n}^n_{x}  = \left( \frac{ \mathbf F^{-T} \mathbf{n}_{  \partial \hat Q ^n} }{ \|  \mathbf F^{-T} \mathbf{n}_{  \partial \hat Q ^n} \| } \right) _x,
\end{equation}
where $\mathbf{n}_{ \partial \hat Q ^n}$ is the normal vector in the undeformed space-time domain.

The integration measures to define the domain change are defined by the jacobians of the deformed domain. The integration measure change of the space-time domain $\tilde Q^n$ is given by $J_{Q^n} = J$, where $J = det(\mathbf F)$. The initial boundary Jacobian is defined as $J_{\Omega^{n}} = 
det(\mathbf F_x(t^{n}))$. In the given case $J_{\Omega^{n}}=1$. Note that for small deformations, where  $\min_\Omega (  J ) \approx 1 $, a simpler approach can be considered by assuming a deformed initial state, e.g., $\Omega^{n}=\pmb{\varphi}^{n-1}_h (t^n) ( \Omega^{n-1})$. In this situation, the evaluation $\hat u_h ^{n-1} (t^{n})$ does not require the change of reference space described in \sect{sec:inter-slab}. This approach is equivalent to a time slab with more than one cell in time and continuous \ac{fe} spaces in time could be considered within this macro-cell.

The pullback of the area differential form to the reference domain is expressed as 
\begin{equation}
da_{\partial \tilde Q^n} = J \sqrt{\mathbf{n} ^T_{\partial \hat Q ^n} \cdot \mathbf C^ {-1} \cdot \mathbf{n} _ {\partial \hat Q ^n} } ~ da_{\partial \hat Q^n}
\end{equation}
where $\mathbf C = \mathbf F^T \mathbf F$ (see, e.g., \cite{Bonet1997}). 

The space-time gradients are transported to the deformed domain,  $\boldsymbol{\nabla} ^n = \mathbf F^{-T} \hat{\boldsymbol{\nabla}}$. The map gradient, $\mathbf F = \boldsymbol{\nabla} \pmb{\varphi}_h^n$,  contains the terms as follows:
\begin{equation}
  \mathbf F = \left[
  \begin{matrix}
    \mathbf F_x & 0 \\
    \partial _t \pmb{\varphi}_{h,x}^n & 1 
  \end{matrix}
  \right]
 =
  \left[
    \begin{matrix}
      \mathbf F_x & 0 \\
      \mathbf w^T & 1 
    \end{matrix}
    \right],
\end{equation}
where $\mathbf{F}_x$ is the space gradient and $\mathbf w^T = \partial _t \pmb{\varphi}_{h,x}^n$ the deformation velocity. Then, the inverse gradient is given by 
\begin{equation}
    \mathbf F^{-T} = \left[
    \begin{matrix}
      \mathbf F_x ^ {-1} & 0 \\
      -\mathbf w^T \mathbf F_x ^{-1} & 1 
    \end{matrix}
    \right].
\end{equation}
Then, by decomposing the space-time gradient into space and time derivatives, we obtain 
  \begin{equation}
    \left[
    \begin{matrix}
        \boldsymbol{\nabla} ^n _x \\
        \partial ^n _t
    \end{matrix}
    \right]
    =
    \mathbf F^{-T}
    \left[
    \begin{matrix}
        \hat{\boldsymbol{\nabla}} _x \\
        \hat \partial _t
    \end{matrix}
    \right]
    =
    \left[
    \begin{matrix}
        \mathbf F^{-1}_x \hat{\boldsymbol{\nabla}} _x \\
        \hat \partial _t - \mathbf w^T  \mathbf F^{-1}_x \hat{\boldsymbol{\nabla}} _x
    \end{matrix}
    \right],
  \end{equation}
which already recovers the derivative terms used in \ac{ale} formulations.

\subsection{Inter-slab integration}\label{sec:inter-slab}
In the formulation \eqref{eq:bilinear-form}-\eqref{eq:linear-form}, a \ac{dg} method is used in time, where the initial value at the time slab is imposed weakly through an inter-time slab jump $ [\hat u^{n} (t^{n}) - \hat u ^{n-1} (t^{n})]$. However, integrating this jump is not straightforward, since $\hat u^{n} (t^{n})$ and $\hat u ^{n-1} (t^{n})$ in \eqref{eq:linear-form} are expressed in different discrete spaces (and meshes).

\begin{figure}[http]
  \centering
  \begin{subfigure}{0.32\textwidth}
    \includegraphics[width=\textwidth]{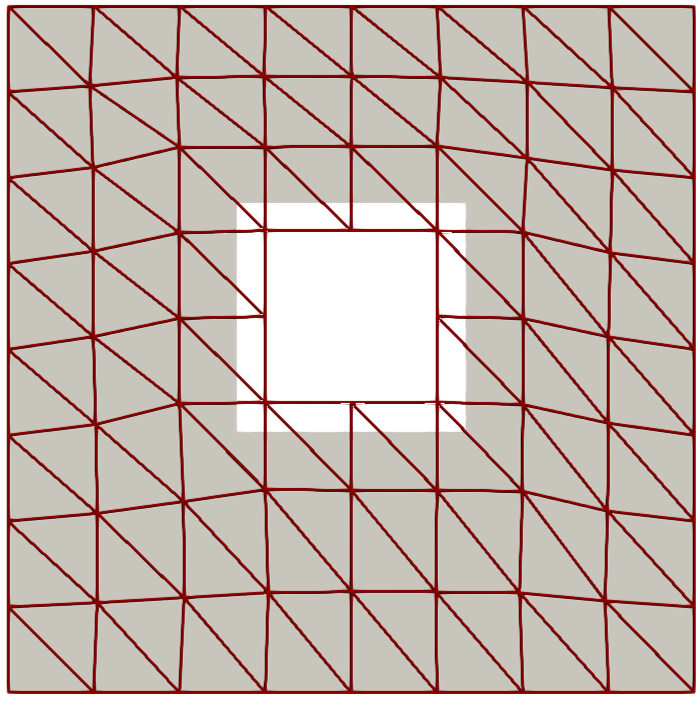}
    \caption{$\pmb{\varphi}^{n-1}_h (t^{n}) (\bar{\mathcal{T}}^{n-1})$}
  \end{subfigure}
  \begin{subfigure}{0.32\textwidth}
    \includegraphics[width=\textwidth]{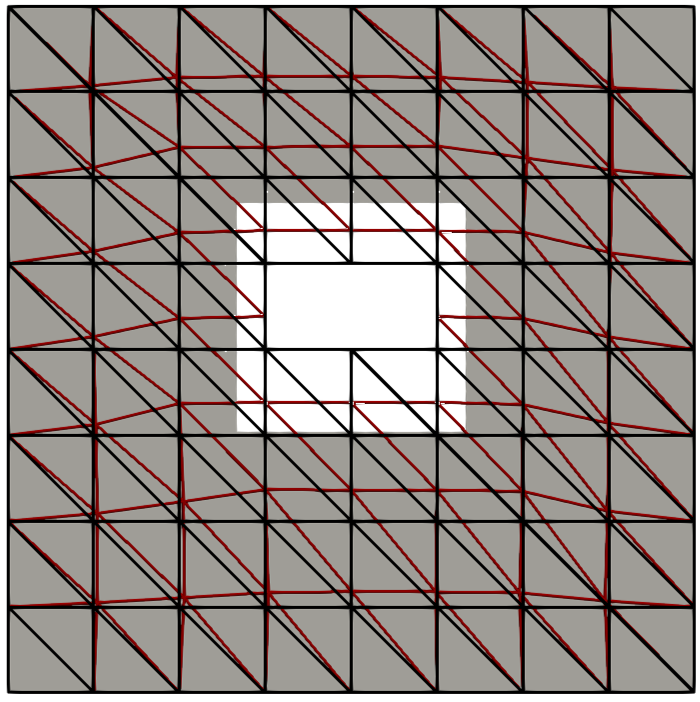}
    \caption{$\bar{\mathcal{T}}_{\mathrm{int}}^{n}$ }
  \end{subfigure}
  \begin{subfigure}{0.32\textwidth}
    \includegraphics[width=\textwidth]{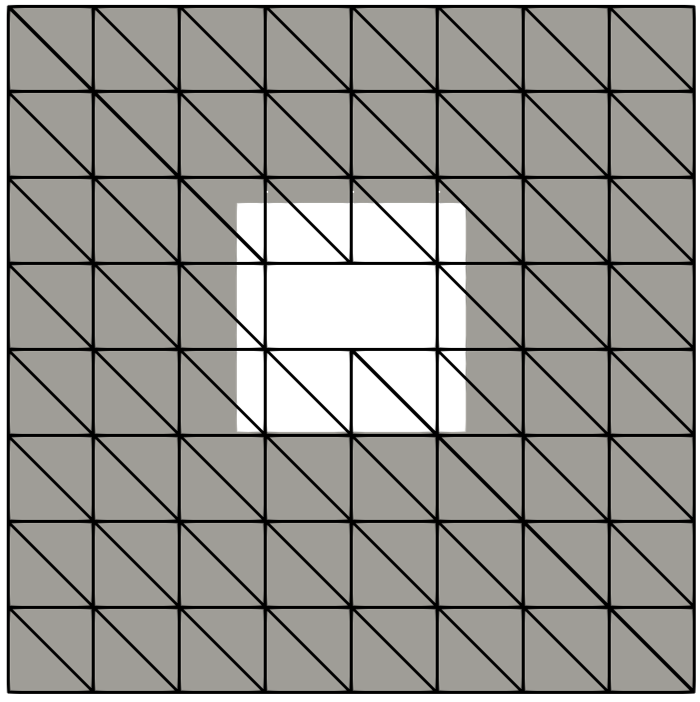}
    \caption{$\bar{\mathcal{T}}_{}^{n}$ }
  \end{subfigure}
  \caption[Mesh sequence for solution transfer between time slabs.]{Mesh sequence for solution transfer between time slabs. The solution obtained in $\bar{\mathcal{T}}_{}^{n-1}$ in (a) is then evaluated at $\bar{\mathcal{T}}_{}^{n}$ (c). Acting as a bridge, the intersected mesh $\bar{\mathcal{T}}_{\mathrm{int}}^{n}= \pmb{\varphi}^{n-1}_h (t^{n}) ( \bar{\mathcal{T}}^{n-1} )\cap  \bar{\mathcal{T}}^{n} \cap \Omega ^{n}$ in (b) facilitates the evaluation by providing injective cell maps to both active meshes. }
  \label{fig:cutter}
\end{figure}

To address this evaluation, {let us introduce a time slab transfer mechanism that relies on an intermediate unfitted discretization $\bar{\mathcal{T}}_{\mathrm{int}}^{n}$}. $\bar{\mathcal{T}}_{\mathrm{int}}^{n}$ is a partition of
$\Omega^{n}$ that results of intersecting the embedded discretization $\mathcal{T}_{\mathrm{cut}}^{n}$ of $\Omega^{n}$ and 
{ $\bar{\mathcal{T}}^{n}_{-} = \pmb{\varphi}^{n-1}_h (t^{n}) ( \bar{\mathcal{T}}^{n-1}_{})$, 
i.e., $\bar{\mathcal{T}}_{\mathrm{int}}^{n}$ results from intersecting $\bar{\mathcal{T}}^n$, $\bar{\mathcal{T}}^n_-$ and $\Omega^{n-1}$. }
Each cell 
$K_\mathrm{int} \in \bar{\mathcal{T}}_{\mathrm{int}}^{n}$
has an injective map to 
${K}^{n-1} \in \bar{\mathcal{T}}^{n-1}_{}$
and 
${K}^{n} \in \bar{\mathcal{T}}^{n}_{}$. 
A representation of $\bar{\mathcal{T}}_{\mathrm{int}}^{n}$ is depicted in \fig{fig:cutter}, and its construction is detailed in \sect{sec:intersection}.

\begin{figure}[http]
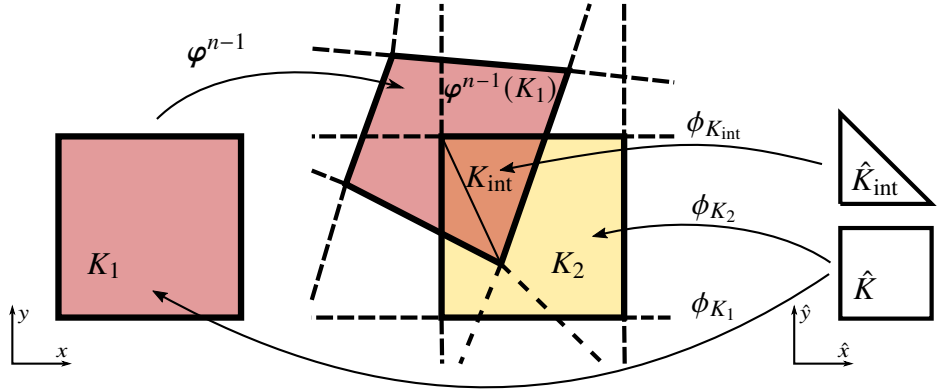

  \centering
  \includefig[\normalsize]{0.8\textwidth}{maps}
  \caption[Mesh sequence for solution transfer between time slabs.]{Representation of the cell maps used in the time slab interface integration. The maps $\phi_{K_1}$ and $\phi_{K_2}$ send the reference element $\hat K$ to the physical space of $K_1\in\bar{\mathcal{T}}^{n-1}$ and $K_2\in\bar{\mathcal{T}}^{n}$, resp. The map $\phi_{K_\mathrm{int}}$ sends the reference element $\hat K_\mathrm{int}$ to the physical space of $K_\mathrm{int}\in\bar{\mathcal{T}}^{n-1}_\mathrm{int}$.}
  \label{fig:cell-maps}
\end{figure}

The integration of the jump in the time slab interface is performed in $\bar{\mathcal{T}}_{\mathrm{int}}^{n}$. Thus, we need to define the cell maps from $\bar{\mathcal{T}}_{\mathrm{int}}^{n}$ to $\bar{\mathcal{T}}^{n}_{}$ and $\bar{\mathcal{T}}^{n+1}_{}$. For $\hat u^{n-1} (t^n)$ we need a cell map from $\hat K_\mathrm{int} \in \bar{\mathcal{T}}_{\mathrm{int}}^{n}$ to $\hat K_1 \in \bar{\mathcal{T}}^{n}_{}$ such that 
\begin{equation}
  \psi^- = \left( \phi_{K_1} \right) ^{-1} \circ \left( \pmb{\varphi}^{n-1} \right) ^{-1} \circ \left( \phi_{K_\mathrm{int}} \right), 
\end{equation}
where $\phi_{K_1}$ maps from the reference space to the undeformed physical space of $K_1 \in \bar{\mathcal{T}}^{n}$
and $\phi_{K_\mathrm{int}}$ maps from the reference space to the physical space of $K_\mathrm{int}$, see \fig{fig:cell-maps}. In the case of $\hat u^{n} (t^{n})$ we need a cell map from $\hat K_\mathrm{int} \in \bar{\mathcal{T}}_{\mathrm{int}}^{n}$ to $\hat K ^{n} \in \bar{\mathcal{T}}^{n}_{}$ such that
\begin{equation}
  \psi^+ = \left( \phi_{K_2} \right) ^{-1}  \circ \left( \phi_{K_\mathrm{int}} \right).
\end{equation}
where $\phi_{K_2}$ maps from the reference space to the physical space of $K_2 \in \bar{\mathcal{T}}^{n}$. Now, we can numerically integrate the jump by evaluating the following integral,
\begin{equation}
  \sum_{K_\mathrm{int}\in\bar{\mathcal{T}}_\mathrm{int}} \sum _ {\hat q} 
  \left(  \hat u^{n} (t^{n}) \circ \psi^+ 
  -  
  \hat u^{n-1} (t^{n}) \circ \psi^- \right) (v^{n}(t^{n}) \circ \psi^+ \left( \hat q \right)) |J_{K_\mathrm{int}}| w_{K_\mathrm{int}},
\end{equation}
where $J_{K_\mathrm{int}}$, $\hat q$ and $w_{K_\mathrm{int}}$ are the Jacobian, quadrature points and quadrature weights of $K_\mathrm{int}$, resp.

\section{Intersection algorithm for time slab transfer}\label{sec:intersection}
In this section, we will present one of the key novelties of this work. We will develop the geometrical algorithms required in \sect{sec:inter-slab} to evaluate the {inter-slab} solution {with the time slab transfer mechanism}.  These algorithms are based on the methods presented in \cite{Badia_2022-stl}. Specifically, we compute a linear intersection  $\bar{\mathcal{T}}_{\mathrm{int}}^{n}$ of $\mathcal{T}_{}^{n}$, $ \bar{\mathcal{T}}^{n}_{-}  = \pmb{\varphi}^{n-1}_h (t^{n}) ( \bar{\mathcal{T}}^{n-1}_{})$ and $\Omega^{n}=\mathrm{int}(\mathcal B_h(t^{n}))$. It is important to note that, to ensure the linearity of the intersections, we need to avoid bilinear terms in $\pmb{\varphi}^{n-1}_h$
using a $P_k$ space, e.g., 
decomposing $K\in \bar{\mathcal{T}}^{n}$ into simplices.

To simplify the exposition of the intersection algorithm \alg{alg:intersection}, we redefine $\mathcal{T} = \bar{\mathcal{T}}_{}^{n}$, ${\mathcal{T}}_- = \bar{\mathcal{T}}_{-}^{n}$ and $\mathcal{B} = \mathcal{B}_h(t^{n})$. Within the cell loop of \alg{alg:intersection}, we first consider the cells close to $K\in\mathcal T$ by restricting $\mathcal{B}$ and $\mathcal{T}_-$ accordingly (\alglin{ln:intersection-1} and \alglin{ln:intersection-2}). These restriction queries are computed during a preprocessing stage before entering the loop. Next, in \alglin{ln:intersection-3}, we compute the intersection of $K$ with the interior of $\mathcal{B}$ using the algorithms in the loop-body of Algorithm 10 in \cite{Badia_2022-stl}. 

These algorithms assume that $\mathcal{B}$ is a linear polytope, which is non-convex in general. Thus, its domain interior $\mathrm{int}(\mathcal B)$ is bounded by the set of planar faces of $\mathcal B$. The intersection $K\cap \mathrm{int}(\mathcal B)$ requires a convex decomposition of $\mathcal B$ and $K$ before intersecting the half-spaces defined by the planar faces (see \cite{Badia_2022-stl}).

Finally, we intersect each polytope in $\mathcal{T}_\mathrm{cut}^K$  by the subset of $\mathcal{T}_-$ around $K$ (\alglin{ln:intersection-4}). These intersections are performed employing convex linear clipping algorithms \cite{Sugihara1994} described in Algorithm 2 in~\cite{Badia_2022-stl}. 
Alternatively, in \alglin{ln:intersection-5}, we intersect the cells $K\in\mathcal{T}$ within the domain $\mathrm{int}(\mathcal{B})$ 
bounded by $\mathcal{B}$ that are not intersected by the domain boundary $\mathcal{B}$.
The information about the cells inside $\mathrm{int}(\mathcal{B})$ is obtained from the propagation through the untouched cells (see \cite{Badia_2022-stl}). The returned triangulation $\mathcal{T}_\mathrm{int}$ not only covers the cells cut cells intersected by the boundary $\mathcal{B}$ but the entire domain enclosed by $\mathcal{B}$. This process guarantees that each cell $K_\mathrm{int} \in {\mathcal{T}}_\mathrm{int}$ has an injective map to $K\in \mathcal{T}$ and $K_- \in \mathcal{T}_-$. {The cells in  ${\mathcal{T}}_\mathrm{int}$ are general polytopes that cannot use standard quadrature rules. In order to numerically integrate in these cells, one can perform a decomposition of the cells into simplices. Alternatively, one can reduce the dimension of the integrals with  Stokes theorem and moment-fitting methods; see more details in \cite{Martorell2023} and references therein.}

\begin{algorithm}
  \caption{$\mathcal{T}\cap\mathcal{T}_-\cap\mathrm{int}(\mathcal{B})$}
    \begin{algorithmic}[1]
    \STATE $\mathcal{T}_\mathrm{cut} \gets \emptyset,\quad \mathcal{T}_\mathrm{in} \gets \emptyset$
    \FOR{$K \in \mathcal{T}$}
      \STATE $\mathcal B^K \gets \mathtt{restrict} (\mathcal{B},K) $ \label{ln:intersection-1}
      \STATE $\mathcal{T}_-^K \gets \mathtt{restrict} (\mathcal{T}_-,K)$\label{ln:intersection-2}
      \STATE $\mathcal{T}_\mathrm{cut}^{K} \gets K \cap \mathrm{int}(\mathcal B^K) $ \label{ln:intersection-3}
      \FOR{$K_- \in \mathcal{T}_-^K$}
        \IF{$K \cap \mathcal{B} \neq \emptyset$}
        \STATE $ \mathcal{T}_\mathrm{cut}^{K,K_-} \gets \{ K_{\mathrm{cut}} \cap K_- : K_{\mathrm{cut}} \in \mathcal{T}_{\mathrm{cut}} ^K \}; \quad \mathcal{T}_\mathrm{cut} \gets \mathcal{T}_\mathrm{cut} \cup \mathcal{T}_\mathrm{cut}^{K,K_-}$ \label{ln:intersection-4}
        \ELSIF{$K \subset \mathrm{int}(\mathcal{B})$}
        \STATE $ \mathcal{T}_\mathrm{in} \gets \mathcal{T}_\mathrm{in} \cup ( K \cap K_-)$ \label{ln:intersection-5}
        \ENDIF
      \ENDFOR
    \ENDFOR
    \RETURN $\mathcal{T}_\mathrm{int} \gets \mathcal{T}_\mathrm{cut} \cap \mathcal{T}_\mathrm{in}$
  \end{algorithmic}
  \label{alg:intersection}
\end{algorithm}

We emphasize that the intersection algorithm is as robust as the core cut algorithm in \alglin{ln:intersection-3}. Furthermore, it is important to note that, for evaluation purposes, accurate tolerance management is not required in the intersection process of \alglin{ln:intersection-4}. While the algorithm is described for the core cut algorithm with an exact embedded discretization of explicit domain representation \cite{Badia_2023}, \alg{alg:intersection} is general enough to be used with other unfitted discretizations.

\section{Numerical experiments}\label{sec:results}
\subsection{Objectives}
In the experiments of this section, we aim to demonstrate the effectiveness of the presented methods. In particular, we examine the following aspects within our formulation:
\begin{itemize}
  \item We evaluate the $hp$-convergence on both 2D and 3D spatial domains, compared to the results and analysis in \cite{Badia_2023}.
  \item We explore numerical stability concerning the cut location and the approximation degrees.
  \item We assess the applicability of our formulation to complex 2D and 3D moving domains derived from \ac{stl} models.
\end{itemize}

It is important to note that these experiments focus on comparing our formulation with the one presented in \cite{Badia_2023}. Recall that \cite{Badia_2023} uses space-time embedded discretizations on implicit geometries determined by level sets, and the results are presented exclusively in 2D+1D domains (as the geometrical algorithms for 3D+1D domains are significantly more complex). In contrast, we design a space-time formulation that works on explicit boundary representations and only require geometrical intersection algorithms in space only, e.g., 2D and 3D vs. 3D and 4D in \cite{Badia_2023}. This is possible by pulling back the problem into an extruded space-time domain using the formulation in Sect. \ref{sec:weak-form}.

\subsection{Environment setup}\label{sec:experiments-setup}

The numerical experiments have been performed on Gadi and Titani supercomputers. Gadi is a high-end supercomputer at the NCI (Australia) with 4962 nodes, 3074 of them
powered by a 2 x 24 core Intel Xeon Platinum 8274 (Cascade Lake) at 3.2 GHz and 192 GB RAM.
Titani is located at UPC (Spain) with 6 nodes, 5 of them powered by a 2 x 12 core Intel Xenon E5-2650L v3 at 1.8 GHz and 256 GB RAM.
 The algorithms presented in this work have been implemented in the Julia programming language \cite{Julia-2017}. 
The unfitted FE computations have been performed using the Julia FE library
\texttt{Gridap.jl} \cite{Verdugo_2022} version 0.17.17 and the extension package for unfitted methods \texttt{GridapEmbedded.jl} version 0.8.1 \cite{GridapEmbedded-jl}. \texttt{STLCutters.jl} version 0.1.6 \cite{Martorell_STLCutters_2021} has been used to compute intersection computations on \ac{stl} geometries.
To mitigate excessive computational costs, the condition numbers are computed in the 1-norm using \texttt{cond()} Julia function.

\subsection{Space-time convergence tests}
\begin{figure}[http]
  \centering
  \includegraphics[width=0.95\textwidth]{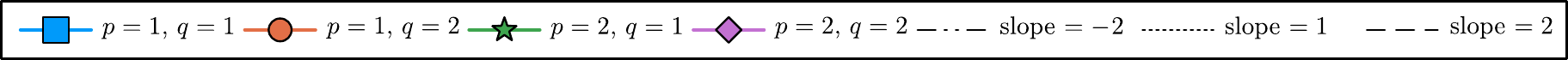}
  \begin{subfigure}{0.24\textwidth}
    \includegraphics[width=\textwidth]{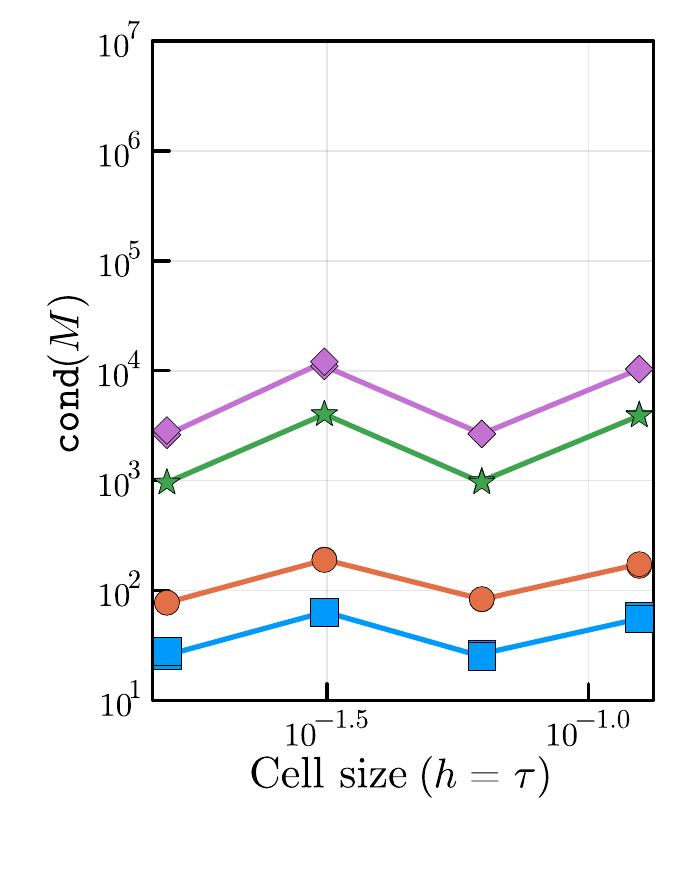}
    \caption{Mass matrix}
    \label{fig:exp1a}
  \end{subfigure}
  \begin{subfigure}{0.24\textwidth}
    \includegraphics[width=\textwidth]{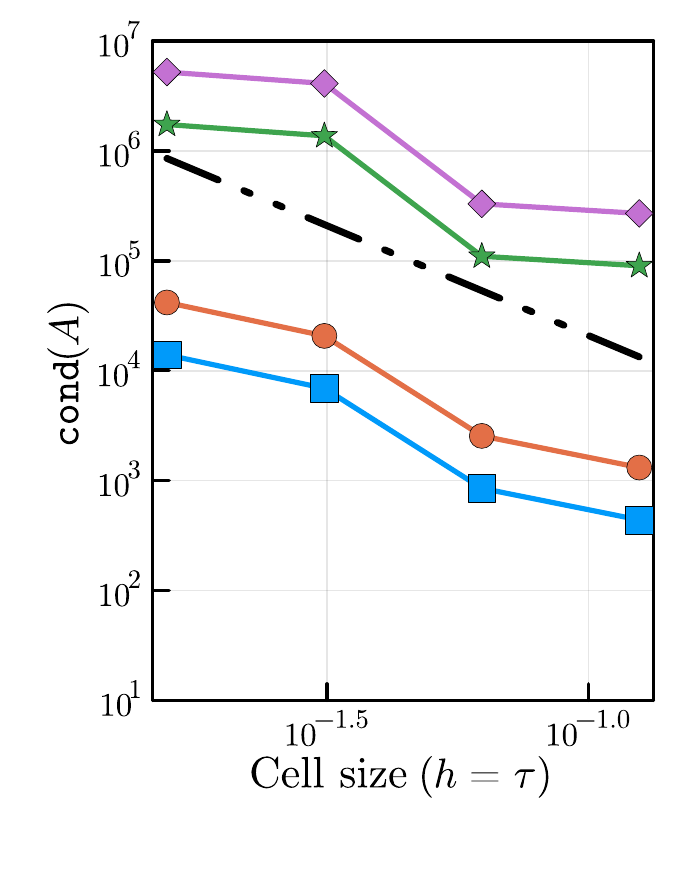}
    \caption{Stiffness matrix}
    \label{fig:exp1b}
  \end{subfigure}
  \begin{subfigure}{0.24\textwidth}
    \includegraphics[width=\textwidth]{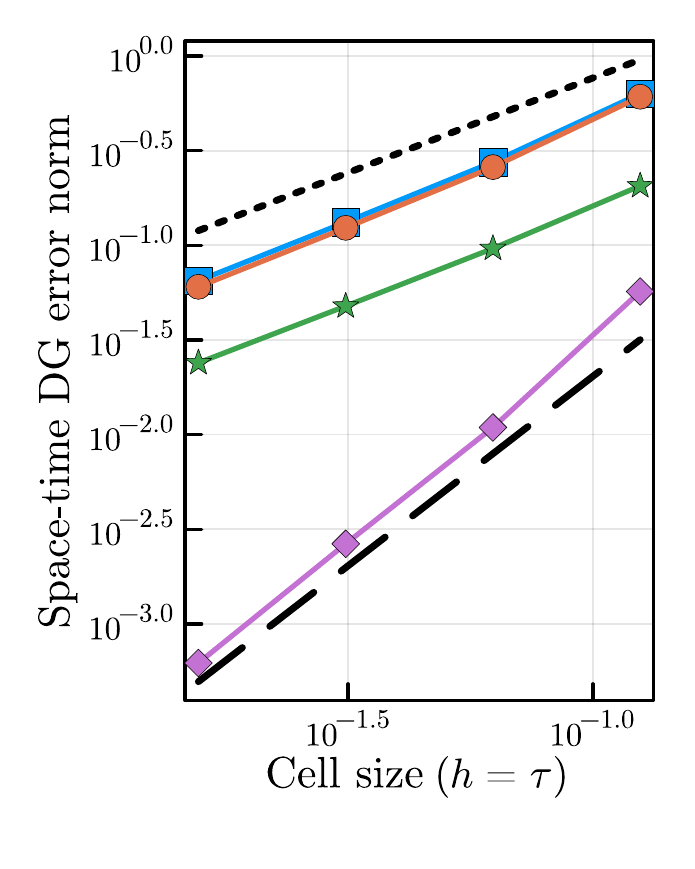}
    \caption{2D}
  \end{subfigure}
  \begin{subfigure}{0.24\textwidth}
    \includegraphics[width=\textwidth]{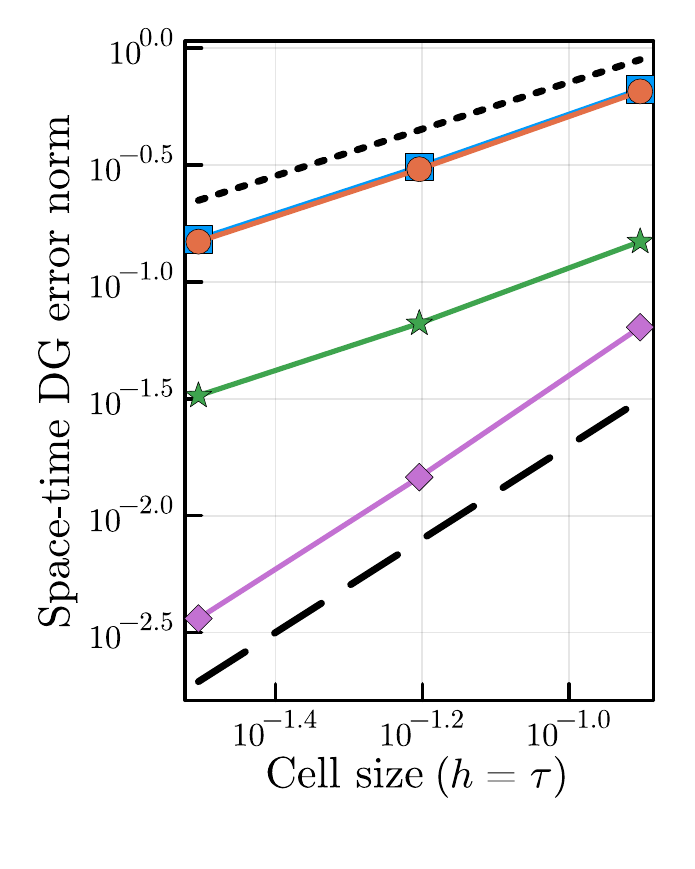}
    \caption{3D}
  \end{subfigure}
  \caption[Scaling of the condition numbers of the mass and stiffness matrices in the initial time slab.  Convergence of the space-time \acs{dg} norm error ($e=u-u_h$) in two and three-dimensional space domains.]{Scaling of the condition numbers of the mass and stiffness matrices in the initial time slab, (a) and (b). Convergence of the space-time \ac{dg} norm error ($e=u-u_h$) in two and three-dimensional space domains (c) and (d). Here, $p$ and $q$ represent the space and time approximation order, resp.}
  \label{fig:exp1}
\end{figure}

\begin{figure}[http]
  \centering
  \includegraphics[width=0.95\textwidth]{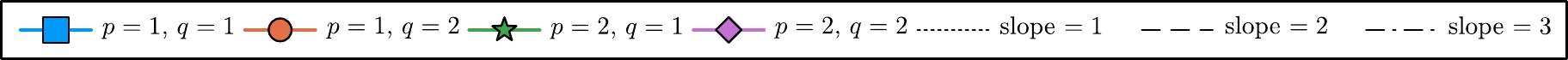}
  \begin{subfigure}{0.24\textwidth}
    \includegraphics[width=\textwidth]{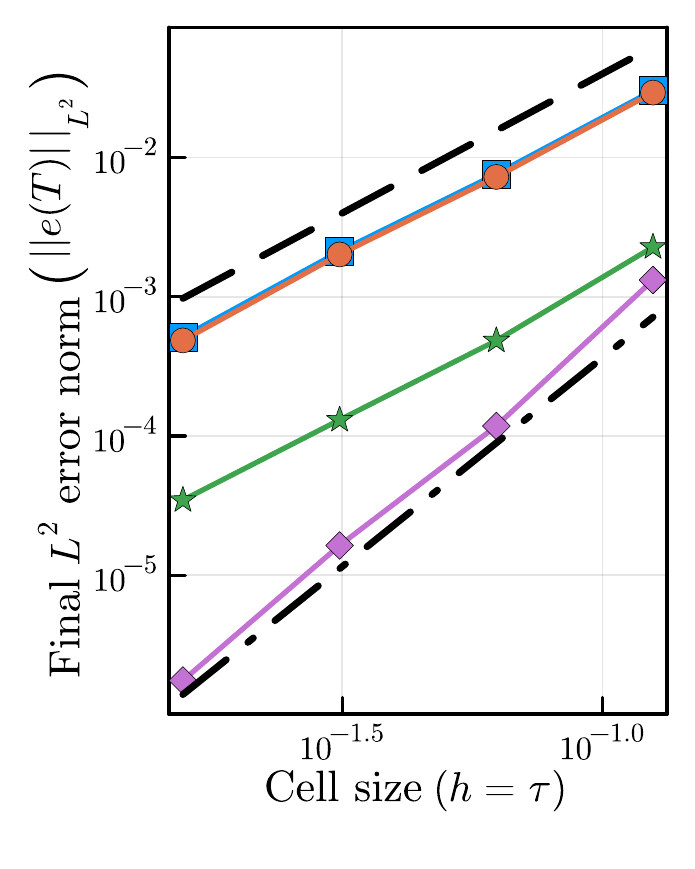}
    \caption{2D}
  \end{subfigure}
  \begin{subfigure}{0.24\textwidth}
    \includegraphics[width=\textwidth]{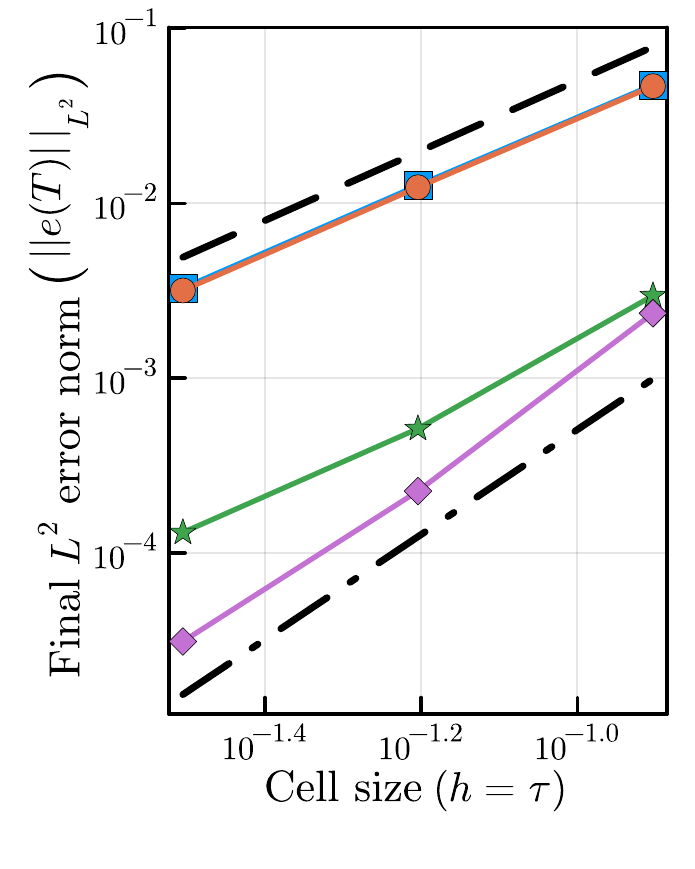}
    \caption{3D}
  \end{subfigure}
  \begin{subfigure}{0.24\textwidth}
    \includegraphics[width=\textwidth]{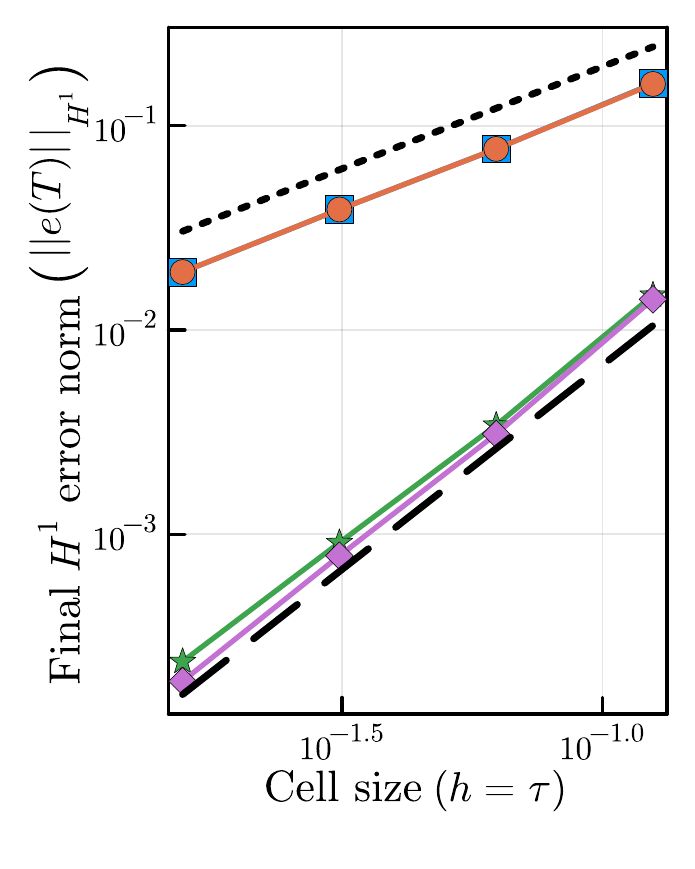}
    \caption{2D}
  \end{subfigure}
  \begin{subfigure}{0.24\textwidth}
    \includegraphics[width=\textwidth]{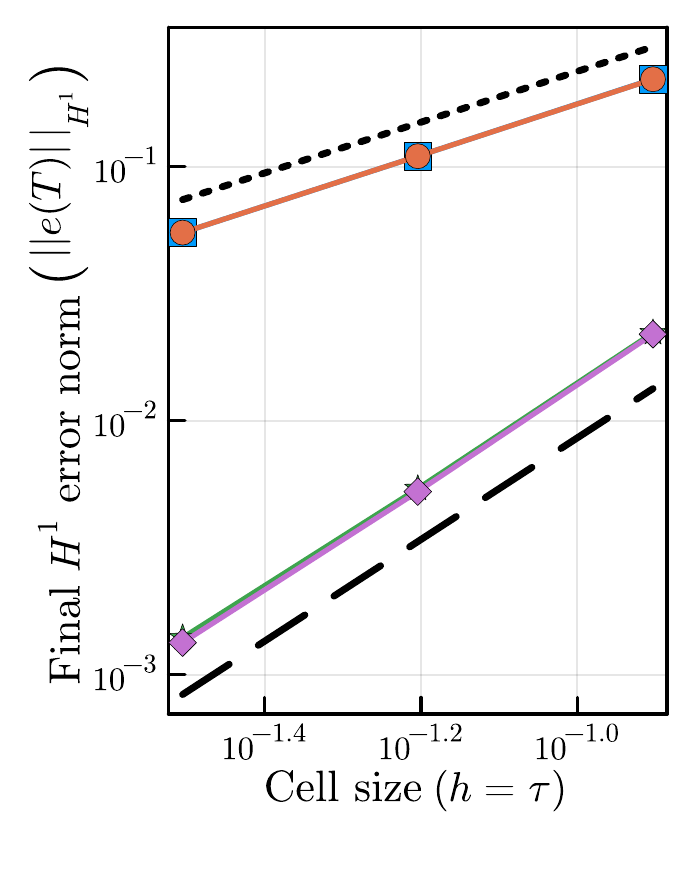}
    \caption{3D}
  \end{subfigure}
  \caption[Convergence of the $L^2(\Omega^n)$ and $H^1(\Omega^n)$ norms of the error ($e=u-u_h$) at the final time $t=T$ in two and three-dimensional space domains.]{Convergence of the $L^2(\Omega^n)$ and $H^1(\Omega^n)$ norms of the error ($e=u-u_h$) at the final time $t=T$ in two and three-dimensional space domains. Here, $p$ and $q$ represent the space and time approximation order, resp.}
  \label{fig:exp2}
\end{figure}

To demonstrate optimal convergence rates we solve a simple \ac{pde} with a manufactured solution out of the \ac{fe} space. Inspired by \cite{Badia_2023}, we solve the Poisson equation \eqref{eq:heat} with the following manufactured solution
\begin{equation}
  u(x,t) = \sin\left( \frac{ \pi \alpha t}{T} \right) \prod_{i=1}^{D} \sin\left( \frac{ \pi  x_i}{L_i} \right) ,
\end{equation}
where $\{L_1,...,L_D,T\}$ are the cartesian dimensions of the space-time domain and the time parameter is set to $\alpha = 0.5$ for the experiments shown in \fig{fig:exp1} and \fig{fig:exp2}. Furthermore, in the equation \eqref{eq:heat} we set the diffusion term as $\mu = 1$. The space domain is a $n$-cube with a $n$-cubic hole in the center. The hole is described by the \ac{stl} of a cube. The lengths at the domain sides are $L=3$ while the lengths of the hole sides are $l=1$. The time domain has size $T=1$. The hole is linearly translated in the $x$ direction. The displacement map is described as $\pmb{D}(x,t) = (0.2t,0,0)$. In all the cases, the domain discretization is a regular Cartesian grid, with the same number of elements $n$ in each direction, both space and time. 

For implementation reasons, the two-dimensional examples are computed with a three-dimensional \ac{stl}. Thus, we build a pseudo-two-dimensional domain
that has only one cell in the $z$ direction. The space-time domain dimensions are $\{L,L,\frac{L}{n},T\}$ while the number of elements in each direction is $\{n,n,1,n\}$. The number of elements per direction in convergence tests is $n=2^i, i=3,...,6$ in two-dimensional runs, while $n=2^i, i=3,...,5$ for three-dimensional runs.

We analyze the condition number of the system to be inverted in each {time slab}. Since the system matrix is nonsymmetric, as suggested in \cite{Badia_2023}, a preconditioner for \ac{dg} in time \cite{Smears_2016} can be considered. The effectiveness of this preconditioner depends on the condition numbers the mass and stiffness matrices, which are defined as follows,
\begin{equation}
  \mathbf{M}_{ab} = \int _{\hat Q^n} \mathcal{E}^n (\mathbf{\Phi}^a)  \mathcal{E}^n (\mathbf{\Phi}^b) |J_{\Omega^n}| {d}\hat{\pmb{x}} {d}t,\quad
  \mathbf{A}_{ab} = a_h\left(\mathcal{E}^n (\mathbf{\Phi}^a),  \mathcal{E}^n (\mathbf{\Phi}^b)\right) .
\end{equation}

The condition numbers presented in \fig{fig:exp1a} and \fig{fig:exp1b} are computed in the initial time slab of the two-dimensional convergence experiments. 
We observe that the condition number of the mass matrix remains nearly constant, while the condition number of the stiffness matrix scales with $\mathcal{O}(h^{-2})$.
These observations align with the behavior expected for \ac{agfem} in space-time domains analyzed in \cite{Badia_2023}.

Now, let us analyze the convergence of the error norms. In \fig{fig:exp1}, we can observe the accumulated \ac{dg} error norm convergence with coercivity constant $c_\mu=1$. We observe that using the \ac{agfem} and constant $h/\tau$ the error converges with $\mathcal{O}(h^r)$, where the convergence rate $r=\min(p,q)$. \fig{fig:exp2} shows the $L^2(\Omega^n)$ and $H^1(\Omega^n)$ norms  at the final time $t=T$. The convergence rate of $L^2(\Omega^n)$ norm is $r = \min(p,q)+1$ while the convergence rate of $H^1(\Omega^n)$ norm is $r = p$. These results are in agreement with the theoretical and analytical space-time \ac{agfem} results in \cite{Badia_2022-highorder}.

\subsection{Moving domains examples}
To demonstrate the applicability of the presented algorithms for the simulation of fluids around moving boundaries, we solve the flow around two rotating geometries in 2D and 3D. We solve the Navier-Stokes and Stokes equations resp. with kinematic viscosity $\nu = 10 ^{-2}$. Regardless of the dimension, we run similar setups. Both share the same discretization order for pressure $p_p=1$, velocity $p_u=2$ and time $q$. We utilize the \ac{agfem} aggregating all cut cells. 

\begin{figure}[http]
  \centering
  \includegraphics[width=0.6\textwidth]{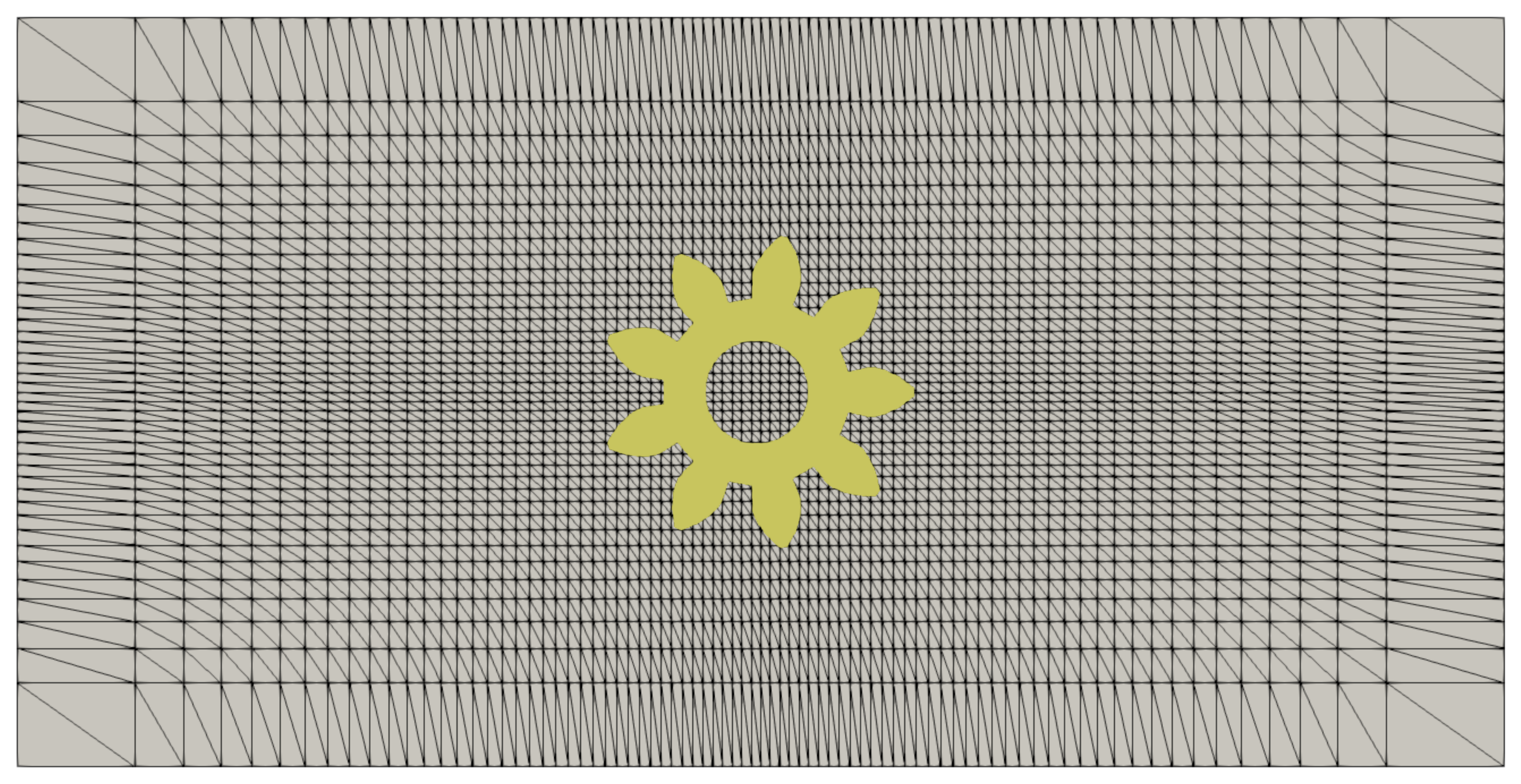}
  \caption[Background spatial mesh $\bar{\mathcal T}^\mathrm{bg}$ around a prismatic gear $\mathcal B_0$.]{Background spatial mesh $\bar{\mathcal T}^\mathrm{bg}$ around a prismatic gear $\mathcal B_0$ (geometry id 71711 from \emph{Thingi10k} \cite{Zhou2016}). This mesh is the simplex partition of a mapped Cartesian mesh. The uniform elements are mapped in each direction with $\phi_M$ \eqref{eq:mesh-map} and an exponential factor $\alpha = 0.5$. The Cartesian mesh has $80\times 40$ elements before the simplex decomposition. The artificial domain $\Omega^\mathrm{art}$ is a box of size $4.8 L_x \times 2.4 L_y$. Here, $L_x$ and $L_y$ represent the bounding box size of $\mathcal B_0$.} 
  \label{fig:mesh-gear}
\end{figure}

\begin{figure}[http]
  \centering
  \includegraphics[width=0.6\textwidth]{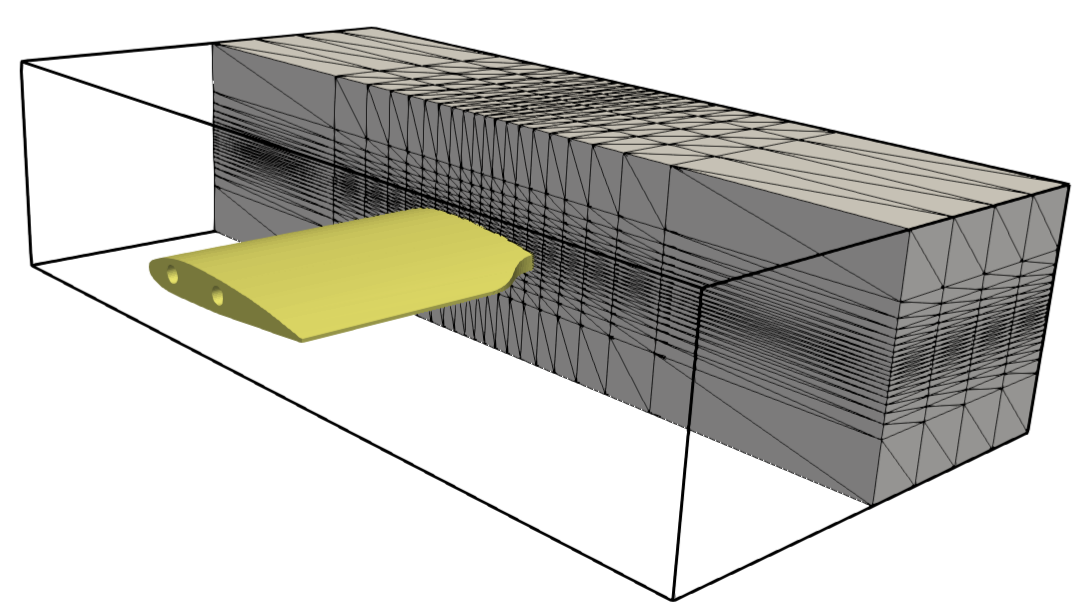}
  \caption[Background spatial mesh $\bar{\mathcal T}^\mathrm{bg}$ around a wing $\mathcal B_0$.]{Background spatial mesh $\bar{\mathcal T}^\mathrm{bg}$ around a wing $\mathcal B_0$ (id 65604 in \emph{Thingi10k}).  
  This simplex mesh comes from a mapped Cartesian mesh. The coordinates of the mesh are mapped in the $x$ and $y$ directions with $\phi_M$ \eqref{eq:mesh-map} and an exponential factor $\alpha = 0.3$. The Cartesian mesh has $20\times 20\times 8$ elements. The artificial domain $\Omega^\mathrm{art}$ is a box of size $4 L_x \times 4 L_y \times 0.8 L_z $. Here, $L_x$, $L_z$ and $L_y$ represent the bounding box size of $\mathcal B_0$.  }
  \label{fig:mesh-wing}
\end{figure}

We improve the accuracy of both meshes by clustering the cells around the geometry. See \fig{fig:mesh-gear} and \fig{fig:mesh-wing} for 2D and 3D, resp. We map each direction $i\in 1,...,d$ of the Cartesian meshes as follows,
\begin{align}\label{eq:mesh-map}
  \phi^i_M(\hat x) = 
  \begin{cases}
    \hat{\pmb{x}}^i _0 \left(\frac{\hat x^i}{\hat x^i_0}\right)^\alpha & \text{if } \hat x^i  < \hat x^i_0, \\
    1 - (1-\hat x^i_0)\left(\frac{1-\hat x^i}{1-\hat x^i_0}\right)^\alpha &\text{otherwise.}
  \end{cases}
\end{align}
Here,  $\hat x^i = x^ i - x^i_0 / L^i$ is the reference axis of the direction $x^i$ where $x_o^i$ is the lower value in the $i$ direction, and $L^i$ is the $i$-length. In \eqref{eq:mesh-map}, 
we use $\hat x^i_0=0.5$ to determine the region of element condensation and the exponential factor $\alpha<1$ for the smoothness of the map.

\begin{figure}[http]
  \centering
  \includefig[\normalsize]{0.3\textwidth}{legend_gear}

  \vspace{0.5cm}

  \begin{subfigure}{0.32\textwidth}
    \includegraphics[width=\textwidth]{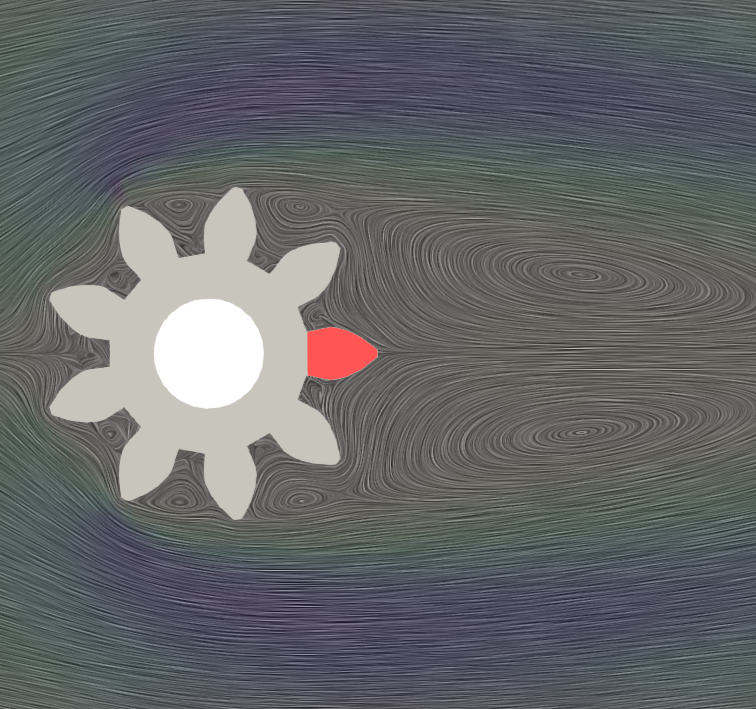}
    \caption{$t=0.0$}
  \end{subfigure}
  \begin{subfigure}{0.32\textwidth}
    \includegraphics[width=\textwidth]{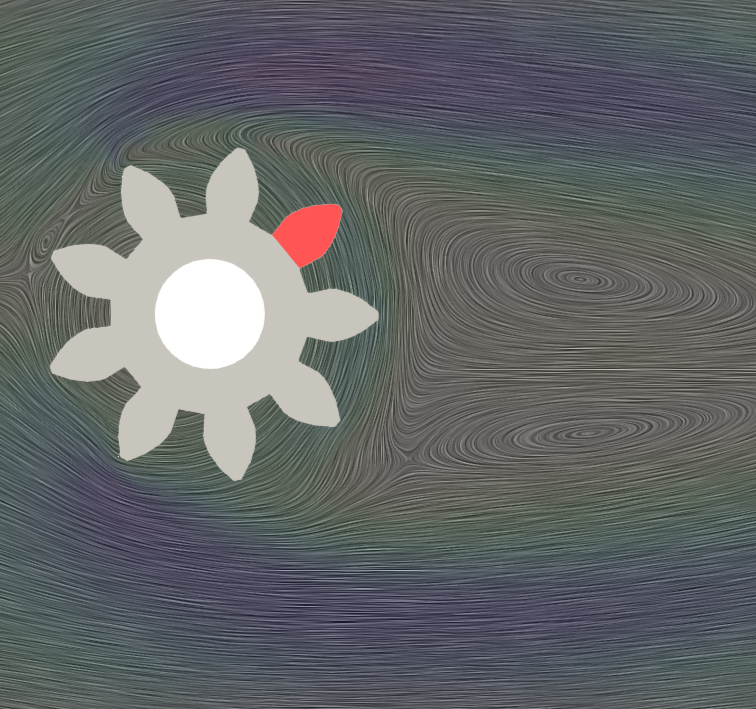}
    \caption{$t=0.5$}
  \end{subfigure}
  \begin{subfigure}{0.32\textwidth}
    \includegraphics[width=\textwidth]{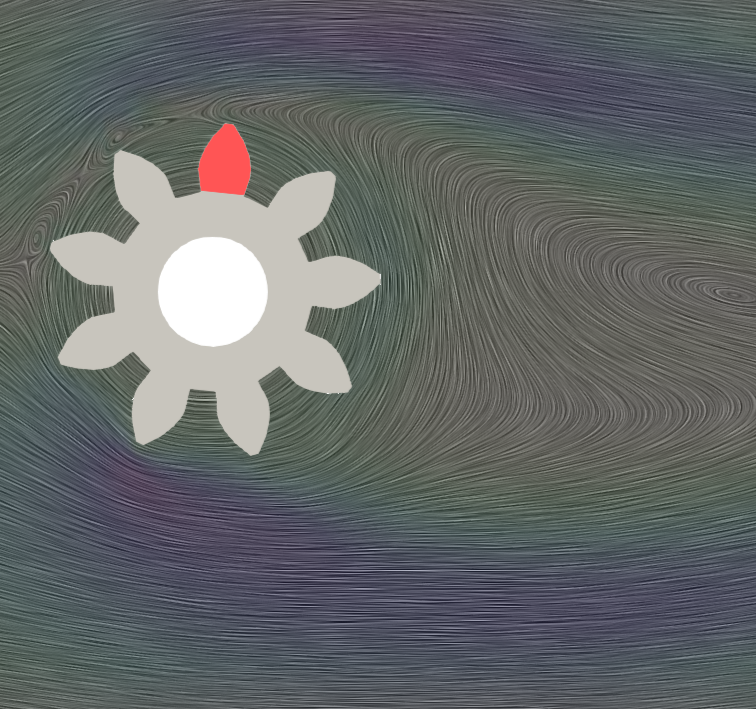}
    \caption{$t=1.0$}
  \end{subfigure}
  \caption[Representation of the \acl{lic} filter of a fluid simulation around an evolving geometry in \fig{fig:spacetime:mesh-gear} at $Re=10^2$.]{ Representation of the \acf{lic} filter of the fluid simulation around an evolving geometry in \fig{fig:mesh-gear} at $Re=10^2$. The time step size is $\tau = 1/60$. The geometry combines rotation and displacements. 
  {The red tooth serves as a reference for visualizing the rotation.}}
  \label{fig:space-time-gear}
\end{figure}

In the 2D example of \fig{fig:mesh-gear} and \fig{fig:space-time-gear}, we solve the Navier-Stokes equations with $Re=10^2$. We define a parabolic inlet flow in the $x$-direction,
\begin{equation}
u(\pmb{x}) = U_\mathrm{max} (4  x_2-4  x_2^2),
\end{equation}
where $U_\mathrm{max} = (1,0,0)$. We set zero velocity on the $y$-faces and zero $z$-velocity in the $z$-faces, i.e., slipping conditions. We weakly impose the displacement velocity $\boldsymbol{\nabla} \pmb{\varphi}^n$ with Nistche's method on the geometry $\mathcal B(t)$.
{
The initial geometry $\mathcal B_0$ is mapped by
$\pmb D = \pmb D_s\circ \pmb{\phi}_t$, \[\pmb D_s(\pmb{x}, t) = \pmb{x}_0 + \mathbf{R}(t) \cdot ( \pmb{x} - \pmb{x}_0) + \mathbf{A}_x \sin(\omega _x t),\] where $\mathbf{R}(t)$ is the rotation matrix with angular velocity $\omega = \pi/2~\mathrm{rad/s}$, $\pmb{x}_0$ is the center of the mesh, $\mathbf{A}_x = (0,0.2)$ and $\omega_x = \pi/2$.
The time map is defined as, $\pmb{\phi}_t(\pmb{x},t) = ( \pmb{x},\phi_t(t))$,
\begin{equation}
  \label{eq:time-map}
  \phi_t = 
  \begin{cases}
    \frac{t_a}{\gamma} \left( \frac{t}{t_a} \right)^\gamma & \text{if } t < t_a, \\
    (t - t_a) + \frac{t_a}{\gamma} & \text{otherwise.}
  \end{cases}
\end{equation}
We set $\gamma=2$ and $t_a=1/8$. With this time map, the initial velocity of the geometry is zero.}

\begin{figure}[http]
  \centering
  \includefig[\normalsize]{0.3\textwidth}{legend_wing}

  \vspace{0.5cm}

  \begin{subfigure}{0.32\textwidth}
    \includegraphics[width=\textwidth]{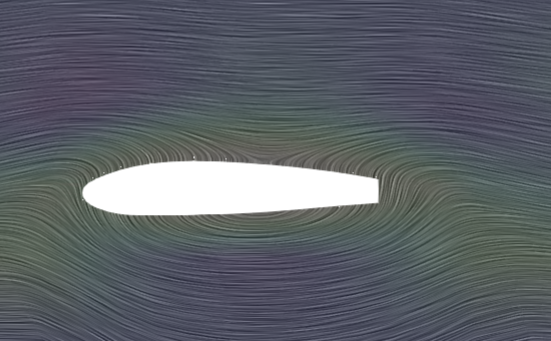}
    \caption{$t=0.0,\, \hat z = 0.5$}
  \end{subfigure}
  \begin{subfigure}{0.32\textwidth}
    \includegraphics[width=\textwidth]{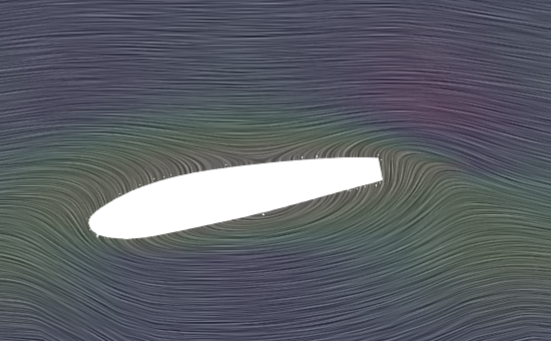}
    \caption{$t=0.1,\, \hat z = 0.5$}
  \end{subfigure}
  \begin{subfigure}{0.32\textwidth}
    \includegraphics[width=\textwidth]{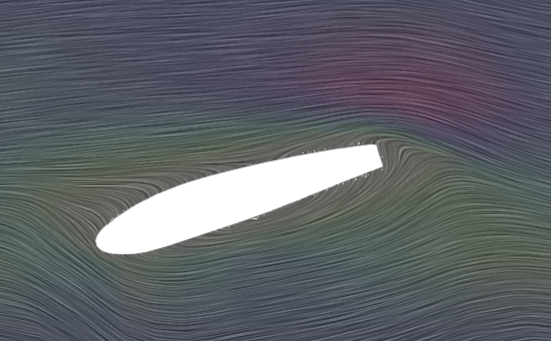}
    \caption{$t=0.2,\, \hat z = 0.5$}
  \end{subfigure}

  \begin{subfigure}{0.32\textwidth}
    \includegraphics[width=\textwidth]{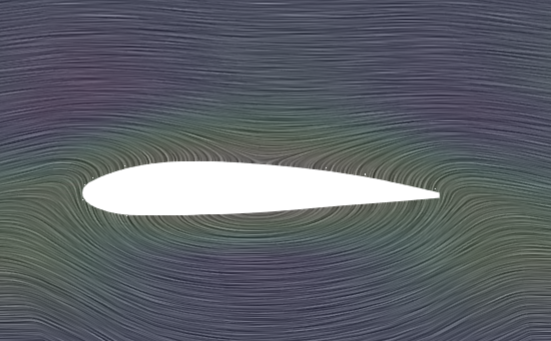}
    \caption{$t=0.0,\, \hat z = 0.75$}
  \end{subfigure}
  \begin{subfigure}{0.32\textwidth}
    \includegraphics[width=\textwidth]{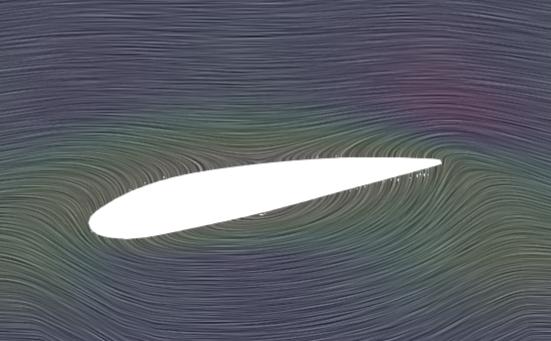}
    \caption{$t=0.1,\, \hat z = 0.75$}
  \end{subfigure}
  \begin{subfigure}{0.32\textwidth}
    \includegraphics[width=\textwidth]{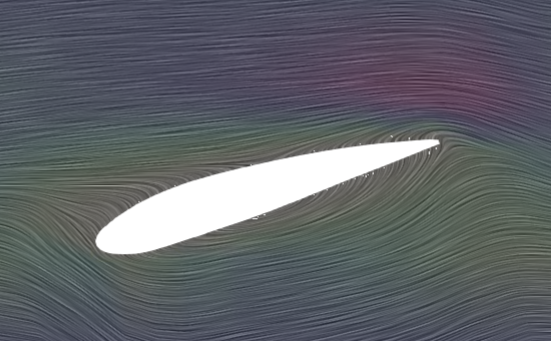}
    \caption{$t=0.2,\, \hat z = 0.75$}
  \end{subfigure}

  \begin{subfigure}{0.32\textwidth}
    \includegraphics[width=\textwidth]{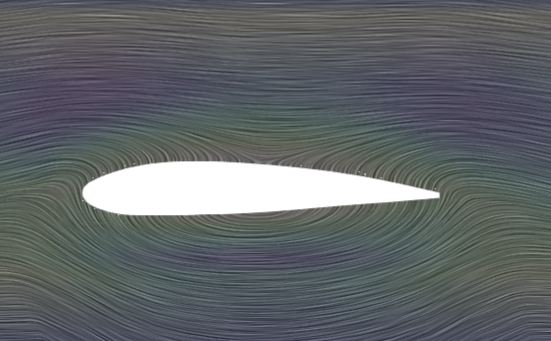}
    \caption{$t=0.0,\, \hat z = 1.0$}
  \end{subfigure}
  \begin{subfigure}{0.32\textwidth}
    \includegraphics[width=\textwidth]{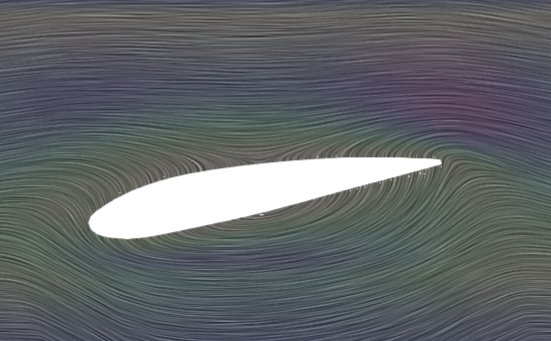}
    \caption{$t=0.1,\, \hat z = 1.0$}
  \end{subfigure}
  \begin{subfigure}{0.32\textwidth}
    \includegraphics[width=\textwidth]{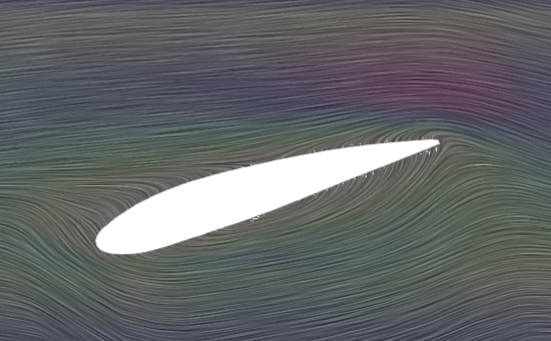}
    \caption{$t=0.2,\, \hat z = 1.0$}
  \end{subfigure}
  \caption[Representation of the \acl{lic} filter of the viscous flow around \fig{fig:spacetime:mesh-wing}.]{Representation of the \ac{lic} filter of the viscous flow around \fig{fig:mesh-wing}. The matrix representation shows several time slabs on different slices. Here, $\hat z=(z-z_0)/L_z$ represents the relative position in $z$ where $z_0$ is the lower the $z$-coordinate and $L_z$ the $z$-length. The step size of the simulation is $\tau = 1/120$. 
  Note that the wing is not an extrusion of an airfoil. The central section ($\hat z = 0.5$) has a flat trailing edge (see \fig{fig:mesh-wing}).
   }
  \label{fig:space-time-wing}
\end{figure}

In the 3D experiment of \fig{fig:mesh-wing} and \fig{fig:space-time-wing}, we simulate a viscous flow ($Re\sim0$) with the same boundary conditions as in the 2D experiment. However, we define a different inlet profile, 
\begin{equation}
  u(\pmb{x}) = U_\mathrm{max} (4  x_2-4  x_2^2) (4  x_3-4  x_3^2),
  \end{equation}
that utilizes the 3D domain. The differences in the $z$-direction are clearly shown in \fig{fig:space-time-wing}. Equally, the $z$-faces have a slipping condition to hold the rotation of the wing in the $z$-direction. 
{
The initial geometry $\mathcal B^0$ is mapped by \[\pmb D (\pmb{x},t) = \pmb{x}_0 + \mathbf{R}_\theta (t) (\pmb{x} - \pmb{x} _ 0),\]
where $\pmb{x}_0$ is the center of the mesh and $\mathbf{R}_\theta$ is the rotation matrix of the angle $\theta (t) = \theta_{\max} \sin( \omega_\theta t)$ over the $z$-axis.
Here, $\theta_{\max} =\pi/10$   and $\omega_\theta = \pi$.
{The 3D experiment is initialized with a non-zero velocity of the geometry.}}

 We slightly optimize the number of time slab transfer operations in these experiments. Above a deformation threshold, e.g., $\max(\boldsymbol{\nabla} \pmb {\varphi}^n) < 0.8$, we maintain the spatial active mesh $\bar{ \mathcal T}^n$ between time slabs. In these cases, the initial value evaluation does not require mesh intersections,  reducing computational cost.

\section{Conclusions and future work}\label{sec:conclusions}

In this work, we have introduced a novel space-time formulation for problems on moving domains with large displacements. The continuous problem at each time slab is pulled back to a constant-in-time domain (obtained by extrusion in time of the initial geometry). The resulting problem is discretized using unfitted \ac{fem}. Intersection algorithms between the boundary of the initial geometry and background mesh are used to compute the required integrals in the formulation. This way, we eliminate the requirement for four-dimensional geometric algorithms. The time discretization relies on \ac{dg} schemes; the jump terms across time slabs are exactly integrated by computing the intersection of the reference and push-forward background meshes.

We have validated this method through an $hp$-convergence analysis of a manufactured solution using \ac{agfem}. We have observed optimal convergence rates for two-dimensional and three-dimensional space domains in these experiments. The convergence has been validated by comparing it with another space-time analysis for unfitted \ac{fem} \cite{Badia_2023}. Additionally, we have observed the expected scaling of the condition number of the mass and stiffness matrices. Furthermore, we have demonstrated the practical applications of this methodology to the simulation of incompressible flows around rotating geometries in two examples, one with a two-dimensional geometry and another with a three-dimensional geometry.

Future work involves the numerical analysis of the method in the case in which the deformation maps do not much the boundary displacement by designing a high-order extension of the work in \cite{Lehrenfeld2019} and the extension of this method to distributed memory machines \cite{Badia_2022-distributed}. The geometrical component of this extension will be highly scalable,
given that the presented algorithms are defined cell-wise and thus embarrasingly parallel, as they are in \cite{Badia_2022-stl}. Distributed computations combined with background mesh refinement, e.g., using octree meshes, will allow us to solve larger real-world applications. Additional developments include applying this method to \ac{fsi} simulations. This extension will develop the full potential of the method in dynamic interface-coupling multiphysics simulations.

\newcommand{\thethanks}{This research was partially funded by the Australian Government through the Australian Research Council (project numbers DP210103092 and DP220103160). We acknowledge Grant PID2021-123611OB-I00 funded by MCIN/AEI/10.13039/501100011033 and by ERDF ``A way of making Europe''. P. A. Martorell acknowledges the support received from Universitat Politècnica de Catalunya and Santander Bank through an FPI fellowship (FPI-UPC 2019). This work was also supported by computational resources provided by the Australian Government through NCI under the National Computational Merit Allocation Scheme.}

\section*{Acknowledgments}

\thethanks  

\setlength{\bibsep}{0.0ex plus 0.00ex} 
\bibliographystyle{myabbrvnat.bst}
\bibliography{refs.bib}

\begin{thebibliography}{44}
\providecommand{\natexlab}[1]{#1}
\providecommand{\url}[1]{\texttt{#1}}
\expandafter\ifx\csname urlstyle\endcsname\relax
  \providecommand{\doi}[1]{doi: #1}\else
  \providecommand{\doi}{doi: \begingroup \urlstyle{rm}\Url}\fi

\bibitem[Karypis et~al.(1997)Karypis, Schloegel, and Kumar]{karypis_parmetis97}
G.~Karypis, K.~Schloegel, and V.~Kumar.
\newblock {P}ar{METIS}: Parallel graph partitioning and sparse matrix ordering
  library.
\newblock Technical report, Department of Computer Science and Engineering,
  University of Minnesota, 1997.

\bibitem[Burstedde et~al.(2011)Burstedde, Wilcox, and
  Ghattas]{burstedde_p4est_2011}
C.~Burstedde, L.~C. Wilcox, and O.~Ghattas.
\newblock p4est: Scalable algorithms for parallel adaptive mesh refinement on
  forests of octrees.
\newblock \emph{{SIAM} Journal on Scientific Computing}, 33\penalty0
  (3):\penalty0 1103--1133, Jan. 2011.
\newblock \doi{10.1137/100791634}.

\bibitem[Burstedde and Holke(2016)]{Burstedde2016}
C.~Burstedde and J.~Holke.
\newblock A tetrahedral space-filling curve for nonconforming adaptive meshes.
\newblock \emph{{SIAM} Journal on Scientific Computing}, 38\penalty0
  (5):\penalty0 C471--C503, Jan. 2016.
\newblock \doi{10.1137/15m1040049}.

\bibitem[Burman and Fern{\'{a}}ndez(2014)]{Burman_2014a}
E.~Burman and M.~A. Fern{\'{a}}ndez.
\newblock An unfitted nitsche method for incompressible
  fluid{\textendash}structure interaction using overlapping meshes.
\newblock \emph{Computer Methods in Applied Mechanics and Engineering},
  279:\penalty0 497--514, sep 2014.
\newblock \doi{10.1016/j.cma.2014.07.007}.

\bibitem[Formaggia et~al.(2021)Formaggia, Gatti, and Zonca]{Formaggia_2021}
L.~Formaggia, F.~Gatti, and S.~Zonca.
\newblock An {XFEM}/{DG} approach for fluid-structure interaction problems with
  contact.
\newblock \emph{Applications of Mathematics}, 66\penalty0 (2):\penalty0
  183--211, jan 2021.
\newblock \doi{10.21136/am.2021.0310-19}.

\bibitem[Schott et~al.(2019)Schott, Ager, and Wall]{Schott_2019}
B.~Schott, C.~Ager, and W.~A. Wall.
\newblock Monolithic cut finite~element{\textendash}based approaches for
  fluid-structure interaction.
\newblock \emph{International Journal for Numerical Methods in Engineering},
  119\penalty0 (8):\penalty0 757--796, apr 2019.
\newblock \doi{10.1002/nme.6072}.

\bibitem[Dekker et~al.(2019)Dekker, Meer, Maljaars, and Sluys]{Dekker_2019}
R.~Dekker, F.~Meer, J.~Maljaars, and L.~Sluys.
\newblock A cohesive {XFEM} model for simulating fatigue crack growth under
  mixed-mode loading and overloading.
\newblock \emph{International Journal for Numerical Methods in Engineering},
  118\penalty0 (10):\penalty0 561--577, feb 2019.
\newblock \doi{10.1002/nme.6026}.

\bibitem[Giovanardi et~al.(2017)Giovanardi, Formaggia, Scotti, and
  Zunino]{Giovanardi_2017}
B.~Giovanardi, L.~Formaggia, A.~Scotti, and P.~Zunino.
\newblock Unfitted {FEM} for modelling the interaction of multiple fractures in
  a poroelastic medium.
\newblock In \emph{Lecture Notes in Computational Science and Engineering},
  pages 331--352. Springer International Publishing, 2017.
\newblock \doi{10.1007/978-3-319-71431-8_11}.

\bibitem[Carraturo et~al.(2020)Carraturo, Jomo, Kollmannsberger, Reali,
  Auricchio, and Rank]{Carraturo_2020}
M.~Carraturo, J.~Jomo, S.~Kollmannsberger, A.~Reali, F.~Auricchio, and E.~Rank.
\newblock Modeling and experimental validation of an immersed thermo-mechanical
  part-scale analysis for laser powder bed fusion processes.
\newblock \emph{Additive Manufacturing}, 36:\penalty0 101498, dec 2020.
\newblock \doi{10.1016/j.addma.2020.101498}.

\bibitem[Neiva et~al.(2020)Neiva, Chiumenti, Cervera, Salsi, Piscopo, Badia,
  Mart{\'{\i}}n, Chen, Lee, and Davies]{Neiva_2020}
E.~Neiva, M.~Chiumenti, M.~Cervera, E.~Salsi, G.~Piscopo, S.~Badia, A.~F.
  Mart{\'{\i}}n, Z.~Chen, C.~Lee, and C.~Davies.
\newblock Numerical modelling of heat transfer and experimental validation in
  powder-bed fusion with the virtual domain approximation.
\newblock \emph{Finite Elements in Analysis and Design}, 168:\penalty0 103343,
  jan 2020.
\newblock \doi{10.1016/j.finel.2019.103343}.

\bibitem[Badia et~al.(2021)Badia, Hampton, and Principe]{Badia_2021_uq}
S.~Badia, J.~Hampton, and J.~Principe.
\newblock Embedded multilevel {Monte} {Carlo} for uncertainty quantification in
  random domains.
\newblock \emph{International Journal for Uncertainty Quantification},
  11\penalty0 (1):\penalty0 119--142, 2021.
\newblock \doi{10.1615/int.j.uncertaintyquantification.2021032984}.

\bibitem[Badia et~al.(2022)Badia, Martorell, and Verdugo]{Badia_2022-stl}
S.~Badia, P.~A. Martorell, and F.~Verdugo.
\newblock Geometrical discretisations for unfitted finite elements on explicit
  boundary representations.
\newblock \emph{Journal of Computational Physics}, 460:\penalty0 111162, jul
  2022.
\newblock \doi{10.1016/j.jcp.2022.111162}.

\bibitem[Martorell and Badia(2023)]{Martorell2023}
P.~A. Martorell and S.~Badia.
\newblock High order unfitted finite element discretizations for explicit
  boundary representations.
\newblock \emph{arXiv}, 2023.
\newblock \doi{10.48550/arXiv.2311.14363}.

\bibitem[de~Prenter et~al.(2017)de~Prenter, Verhoosel, van Zwieten, and van
  Brummelen]{DePrenter2017}
F.~de~Prenter, C.~Verhoosel, G.~van Zwieten, and E.~van Brummelen.
\newblock Condition number analysis and preconditioning of the finite cell
  method.
\newblock \emph{Computer Methods in Applied Mechanics and Engineering},
  316:\penalty0 297--327, apr 2017.
\newblock \doi{10.1016/j.cma.2016.07.006}.

\bibitem[Burman(2010)]{burman2010ghost}
E.~Burman.
\newblock Ghost penalty.
\newblock \emph{Comptes Rendus Mathematique}, 348\penalty0 (21-22):\penalty0
  1217--1220, nov 2010.
\newblock \doi{10.1016/j.crma.2010.10.006}.

\bibitem[Burman et~al.(2014)Burman, Claus, Hansbo, Larson, and
  Massing]{burman_cutfem_2015}
E.~Burman, S.~Claus, P.~Hansbo, M.~G. Larson, and A.~Massing.
\newblock {CutFEM}: Discretizing geometry and partial differential equations.
\newblock \emph{International Journal for Numerical Methods in Engineering},
  104\penalty0 (7):\penalty0 472--501, dec 2014.
\newblock \doi{10.1002/nme.4823}.

\bibitem[M{\"u}ller et~al.(2016)M{\"u}ller, Kr{\"a}mer-Eis, Kummer, and
  Oberlack]{muller2017high}
B.~M{\"u}ller, S.~Kr{\"a}mer-Eis, F.~Kummer, and M.~Oberlack.
\newblock A high-order discontinuous galerkin method for compressible flows
  with immersed boundaries.
\newblock \emph{International Journal for Numerical Methods in Engineering},
  110\penalty0 (1):\penalty0 3--30, nov 2016.
\newblock \doi{10.1002/nme.5343}.

\bibitem[Badia et~al.(2018{\natexlab{a}})Badia, Verdugo, and
  Mart{\'{\i}}n]{Badia2018c}
S.~Badia, F.~Verdugo, and A.~F. Mart{\'{\i}}n.
\newblock The aggregated unfitted finite element method for elliptic problems.
\newblock \emph{Computer Methods in Applied Mechanics and Engineering},
  336:\penalty0 533--553, jul 2018{\natexlab{a}}.
\newblock \doi{10.1016/j.cma.2018.03.022}.

\bibitem[Badia et~al.(2018{\natexlab{b}})Badia, Martin, and
  Verdugo]{Badia2018a}
S.~Badia, A.~F. Martin, and F.~Verdugo.
\newblock Mixed aggregated finite element methods for the unfitted
  discretization of the stokes problem.
\newblock \emph{{SIAM} Journal on Scientific Computing}, 40\penalty0
  (6):\penalty0 B1541--B1576, jan 2018{\natexlab{b}}.
\newblock \doi{10.1137/18m1185624}.

\bibitem[Verdugo et~al.(2019)Verdugo, Mart{\'{\i}}n, and Badia]{Verdugo2019}
F.~Verdugo, A.~F. Mart{\'{\i}}n, and S.~Badia.
\newblock Distributed-memory parallelization of the aggregated unfitted finite
  element method.
\newblock \emph{Computer Methods in Applied Mechanics and Engineering},
  357:\penalty0 112583, dec 2019.
\newblock \doi{10.1016/j.cma.2019.112583}.

\bibitem[Badia et~al.(2021)Badia, Mart{\'{\i}}n, Neiva, and
  Verdugo]{Badia_2021a}
S.~Badia, A.~F. Mart{\'{\i}}n, E.~Neiva, and F.~Verdugo.
\newblock The aggregated unfitted finite element method on parallel tree-based
  adaptive meshes.
\newblock \emph{{SIAM} Journal on Scientific Computing}, 43\penalty0
  (3):\penalty0 C203--C234, jan 2021.
\newblock \doi{10.1137/20m1344512}.

\bibitem[Neiva and Badia(2021)]{Neiva2021}
E.~Neiva and S.~Badia.
\newblock Robust and scalable h-adaptive aggregated unfitted finite elements
  for interface elliptic problems.
\newblock \emph{Computer Methods in Applied Mechanics and Engineering},
  380:\penalty0 113769, jul 2021.
\newblock \doi{10.1016/j.cma.2021.113769}.

\bibitem[Badia et~al.(2022{\natexlab{a}})Badia, Neiva, and
  Verdugo]{Badia_2022-highorder}
S.~Badia, E.~Neiva, and F.~Verdugo.
\newblock Robust high-order unfitted finite elements by interpolation-based
  discrete extension.
\newblock \emph{Computers {\&} Mathematics with Applications}, 127:\penalty0
  105--126, dec 2022{\natexlab{a}}.
\newblock \doi{10.1016/j.camwa.2022.09.027}.

\bibitem[Badia et~al.(2022{\natexlab{b}})Badia, Neiva, and
  Verdugo]{Badia_2022-ghost}
S.~Badia, E.~Neiva, and F.~Verdugo.
\newblock Linking ghost penalty and aggregated unfitted methods.
\newblock \emph{Computer Methods in Applied Mechanics and Engineering},
  388:\penalty0 114232, jan 2022{\natexlab{b}}.
\newblock \doi{10.1016/j.cma.2021.114232}.

\bibitem[Beau et~al.(1993)Beau, Ray, Aliabadi, and Tezduyar]{Le_Beau_1993}
G.~L. Beau, S.~Ray, S.~Aliabadi, and T.~Tezduyar.
\newblock {SUPG} finite element computation of compressible flows with the
  entropy and conservation variables formulations.
\newblock \emph{Computer Methods in Applied Mechanics and Engineering},
  104\penalty0 (3):\penalty0 397--422, may 1993.
\newblock \doi{10.1016/0045-7825(93)90033-t}.

\bibitem[Tezduyar et~al.(2006)Tezduyar, Sathe, Keedy, and Stein]{Tezduyar_2006}
T.~E. Tezduyar, S.~Sathe, R.~Keedy, and K.~Stein.
\newblock Space{\textendash}time finite element techniques for computation of
  fluid{\textendash}structure interactions.
\newblock \emph{Computer Methods in Applied Mechanics and Engineering},
  195\penalty0 (17-18):\penalty0 2002--2027, mar 2006.
\newblock \doi{10.1016/j.cma.2004.09.014}.

\bibitem[Thompson and Pinsky(1996)]{Thompson_1996}
L.~L. Thompson and P.~M. Pinsky.
\newblock A space-time finite element method for structural acoustics in
  infinite domains part 1: Formulation, stability and convergence.
\newblock \emph{Computer Methods in Applied Mechanics and Engineering},
  132\penalty0 (3-4):\penalty0 195--227, jun 1996.
\newblock \doi{10.1016/0045-7825(95)00955-8}.

\bibitem[Donea et~al.(1982)Donea, Giuliani, and Halleux]{Donea_1982}
J.~Donea, S.~Giuliani, and J.~Halleux.
\newblock An arbitrary lagrangian-eulerian finite element method for transient
  dynamic fluid-structure interactions.
\newblock \emph{Computer Methods in Applied Mechanics and Engineering},
  33\penalty0 (1-3):\penalty0 689--723, sep 1982.
\newblock \doi{10.1016/0045-7825(82)90128-1}.

\bibitem[Nobile and Formaggia(1999)]{Nobile1999}
F.~Nobile and L.~Formaggia.
\newblock A stability analysis for the arbitrary lagrangian eulerian
  formulation with finite elements.
\newblock \emph{East-West Journal of Numerical Mathematics}, 7\penalty0
  (2):\penalty0 105--132, 1999.

\bibitem[Badia et~al.(2023)Badia, Dilip, and Verdugo]{Badia_2023}
S.~Badia, H.~Dilip, and F.~Verdugo.
\newblock Space-time unfitted finite element methods for time-dependent
  problems on moving domains.
\newblock \emph{Computers {\&} Mathematics with Applications}, 135:\penalty0
  60--76, apr 2023.
\newblock \doi{10.1016/j.camwa.2023.01.032}.

\bibitem[Heimann et~al.(2023)Heimann, Lehrenfeld, and Preuß]{Heimann2023}
F.~Heimann, C.~Lehrenfeld, and J.~Preuß.
\newblock Geometrically higher order unfitted space-time methods for pdes on
  moving domains.
\newblock \emph{SIAM Journal on Scientific Computing}, 45\penalty0
  (2):\penalty0 B139–B165, Mar. 2023.
\newblock \doi{10.1137/22m1476034}.

\bibitem[Lehrenfeld and Olshanskii(2019)]{Lehrenfeld2019}
C.~Lehrenfeld and M.~Olshanskii.
\newblock An eulerian finite element method for pdes in time-dependent domains.
\newblock \emph{ESAIM: Mathematical Modelling and Numerical Analysis},
  53\penalty0 (2):\penalty0 585–614, Mar. 2019.
\newblock \doi{10.1051/m2an/2018068}.

\bibitem[de~Prenter et~al.(2023)de~Prenter, Verhoosel, van Brummelen, Larson,
  and Badia]{dePrenter2023}
F.~de~Prenter, C.~V. Verhoosel, E.~H. van Brummelen, M.~G. Larson, and
  S.~Badia.
\newblock Stability and conditioning of immersed finite element methods:
  Analysis and remedies.
\newblock \emph{Archives of Computational Methods in Engineering}, 30\penalty0
  (6):\penalty0 3617–3656, May 2023.
\newblock \doi{10.1007/s11831-023-09913-0}.

\bibitem[Nitsche(1971)]{Nitsche1971}
J.~Nitsche.
\newblock \"{U}ber ein variationsprinzip zur l\"{o}sung von dirichlet-problemen
  bei verwendung von teilr\"{a}umen, die keinen randbedingungen unterworfen
  sind.
\newblock \emph{Abhandlungen aus dem Mathematischen Seminar der Universit\"{a}t
  Hamburg}, 36\penalty0 (1):\penalty0 9–15, July 1971.
\newblock \doi{10.1007/bf02995904}.

\bibitem[Heimann and Lehrenfeld(2023)]{heimann2023geometrically}
F.~Heimann and C.~Lehrenfeld.
\newblock Geometrically higher order unfitted space-time methods for pdes on
  moving domains: Geometry error analysis.
\newblock \emph{arXiv}, 2023.
\newblock \doi{10.48550/arXiv.2311.02348}.

\bibitem[Bonet and Wood(1997)]{Bonet1997}
J.~Bonet and R.~D. Wood.
\newblock \emph{Nonlinear continuum mechanics for finite element analysis}.
\newblock Cambridge university press, 1997.

\bibitem[Sugihara(1994)]{Sugihara1994}
K.~Sugihara.
\newblock A robust and consistent algorithm for intersecting convex polyhedra.
\newblock \emph{Computer Graphics Forum}, 13\penalty0 (3):\penalty0 45--54, aug
  1994.
\newblock \doi{10.1111/1467-8659.1330045}.

\bibitem[Bezanson et~al.(2017)Bezanson, Edelman, Karpinski, and
  Shah]{Julia-2017}
J.~Bezanson, A.~Edelman, S.~Karpinski, and V.~B. Shah.
\newblock Julia: A fresh approach to numerical computing.
\newblock \emph{{SIAM} Review}, 59\penalty0 (1):\penalty0 65--98, jan 2017.
\newblock \doi{10.1137/141000671}.

\bibitem[Verdugo and Badia(2022)]{Verdugo_2022}
F.~Verdugo and S.~Badia.
\newblock The software design of gridap: A finite element package based on the
  julia {JIT} compiler.
\newblock \emph{Computer Physics Communications}, 276:\penalty0 108341, jul
  2022.
\newblock \doi{10.1016/j.cpc.2022.108341}.

\bibitem[Verdugo et~al.(2023)Verdugo, Neiva, and Badia]{GridapEmbedded-jl}
F.~Verdugo, E.~Neiva, and S.~Badia.
\newblock {GridapEmbedded. Version 0.8.}, Jan. 2023.
\newblock Available at \url{https://github.com/gridap/GridapEmbedded.jl}.

\bibitem[Martorell et~al.(2021)Martorell, Badia, and
  Verdugo]{Martorell_STLCutters_2021}
P.~A. Martorell, S.~Badia, and F.~Verdugo.
\newblock {STLCutters}.
\newblock \emph{Zenodo}, Sept. 2021.
\newblock \doi{10.5281/zenodo.5444427}.

\bibitem[Smears(2016)]{Smears_2016}
I.~Smears.
\newblock Robust and efficient preconditioners for the discontinuous galerkin
  time-stepping method.
\newblock \emph{{IMA} Journal of Numerical Analysis}, page drw050, oct 2016.
\newblock \doi{10.1093/imanum/drw050}.

\bibitem[Zhou and Jacobson(2016)]{Zhou2016}
Q.~Zhou and A.~Jacobson.
\newblock {Thingi10K: A Dataset of 10,000 3D-Printing Models}.
\newblock \emph{arXiv}, 2016.
\newblock \doi{10.48550/arXiv.1605.04797}.

\bibitem[Badia et~al.(2022)Badia, Mart{\'{\i}}n, and
  Verdugo]{Badia_2022-distributed}
S.~Badia, A.~F. Mart{\'{\i}}n, and F.~Verdugo.
\newblock {GridapDistributed}: a massively parallel finite element toolbox in
  julia.
\newblock \emph{Journal of Open Source Software}, 7\penalty0 (74):\penalty0
  4157, jun 2022.
\newblock \doi{10.21105/joss.04157}.

\end{thebibliography}

\end{document}